%% file: prd.tex
\documentclass[twocolumn,showpacs,aps,floatfix,prd,nofootinbib]{revtex4}
\usepackage{graphicx}
\usepackage{dcolumn}
\usepackage{array, longtable}
\usepackage{epsfig}
\usepackage{amsmath}
\usepackage{colordvi}
\usepackage{hhline}

\newcommand{\BaBarYear}{2007}
\newcommand{\BaBarNumber}{003}
\newcommand{\SLACPubNumber}{12377}

 \newcommand{\BaBarType}      {PUB}  % Journal publication
\input babarsym
\input Definitions
\long\def\inst#1{\par\nobreak\kern 4pt\nobreak
    {\it #1}\par\vskip 10pt plus 3pt minus 3pt}

\begin{document}

%

%\begin{flushleft}
%\babar\ Analysis Document \# \BaBarNumber,
%Version 11 \\
%\today \\
%\end{flushleft}

\begin{flushleft}
hep-ex/0703008\\
\babar-\BaBarType-\BaBarYear/\BaBarNumber \\
SLAC-PUB-\SLACPubNumber \\
\end{flushleft}

%\par\vskip 3cm

% Title of the paper
\title{\large \bf
\boldmath
Measurement of {\em CP}-Violating Asymmetries in
\boldmath$B^0\to(\rho\pi)^0$ \\%[0.05cm]
Using a Time-Dependent Dalitz Plot Analysis
} % end title

% Input author list file ?
\input pubboard/authors_jan2007

\date{\today}

% Abstract
\begin{abstract}
We report a measurement of \CP-violating asymmetries in
$B^0\to(\rho\pi)^0\to\pi^+\pi^-\pi^0$ decays using a time-dependent
Dalitz plot analysis. The results are obtained from a data sample
of 375 million $\FourS \to B\Bbar$ decays, collected by the \babar\
detector at the \pep2\ asymmetric-energy \B~Factory at SLAC. 
We measure 26 coefficients of the bilinear form-factor terms occurring
in the time-dependent decay rate of the \Bz meson. 
We derive the physically relevant quantities
from these coefficients. 
In particular, we measure a constraint on the angle $\alpha$
of the Unitarity Triangle.
\end{abstract}

\pacs{13.66.Bc, 14.40.Cs, 13.25.Gv, 13.25.Jx, 13.20.Jf}% PACS

%\vfill
%\centerline{To be Submitted to Physical Review D}
%\newpage
\maketitle

% reset footnote counter
\setcounter{footnote}{0}

% The body of the paper starts here

\input Introduction

\input DetectorAndData

\input AnalysisMethod

\input FitResults

\input Systematics
\input PhysicsResults
\input Summary

\section{Acknowledgments}
\label{sec:acknowledgments}
\input pubboard/acknowledgements

\input Bibliography
\end{document}

%% file: Definitions.tex
\newcommand{\bei}{\begin{itemize}}
\newcommand{\eei}{\end{itemize}}
\newcommand{\beq}{\begin{equation}}
\newcommand{\eeq}{\end{equation}}
\newcommand{\beqn}{\begin{eqnarray}}
\newcommand{\eeqn}{\end{eqnarray}}
\newcommand{\beqns}{\begin{eqnarray*}}
\newcommand{\eeqns}{\end{eqnarray*}}

\newcommand{\intl}{\int\limits}

\newcommand{\e}{\varepsilon}

\def\bk{\!\!\!\!}

\def\ea{{\em et al.}}
\def\min{{\rm min}}

\def\rPTbarkappa{\kern 0.18em\overline{\kern -0.18em r}{}^{\kappa}{}}
\def\rPTbarsigma{\kern 0.18em\overline{\kern -0.18em r}{}^{\sigma}{}}

\def\deltabarkappa{\kern 0.18em\overline{\kern -0.18em \delta}{}_r^{\kappa}}
\def\deltabarsigma{\kern 0.18em\overline{\kern -0.18em \delta}{}_r^{\sigma}}
\def\deltaTbarkappa{\kern 0.18em\overline{\kern -0.18em \delta}{}_T^{\kappa}}
\def\deltaTbarsigma{\kern 0.18em\overline{\kern -0.18em \delta}{}_T^{\sigma}}

\newcommand\ph{\phantom}
\newcommand{\half}{\ensuremath{{1\over2}}}
\newcommand{\pvec}{{\bf p}}

\def\OC{X}
\def\OCbar{{\kern 0.18em\overline{\kern -0.18em \OC}}}
\def\mBz{m_{\Bz}}
\def\spz{s_{+}}
\def\smz{s_{-}}
\def\spm{s_{0}}

\def\mpm{m_{0}}

\def\fpz{f_{+}}
\def\fmz{f_{-}}
\def\fpm{f_{0}}

\def\mpmMax{\mpm^{\rm max}}
\def\mpmMin{\mpm^{\rm min}}

\def\pipipi{\pip\pim\piz}
\def\Btopipipi{\Bz\to\pipipi}

\def\mprime{m^\prime}
\def\thetaprime{\theta^\prime}
\def\deprime{{\de^\prime}{}}
\def\dt{\deltat}

\def\de{\DeltaE}
\def\demin{\de_{-}}
\def\demax{\de_{+}}
\def\deminmax{\de_{\pm}}

\def\tpi{3\pi}

\def\Qtag{q_{\rm tag}}
\def\Qtagi{q_{{\rm tag},i}}
\def\Atagqq{A_{q\bar q,\,\rm tag}}
\def\Atag{A_{{B^+,\,\rm tag}}}
\def\Atagj{A_{{B^+,\,\rm tag},j}}

\def\cat{c}

\def\detJ{|\det J|}
\def\detJi{|\det J_i|}
\def\Rscf{R_{\rm SCF}}

\def\ampfrac{a}
\def\a{\kappa}

\def\c{\tau}

\def\Amptp{{A}_{3\pi}}
\def\Amptpbar{\kern 0.18em\overline{\kern -0.18em {A}}_{3\pi}}

\def\absAmptp{|\Amptp|}
\def\absAmptpbar{|\Amptpbar|}
\def\Amptpkappa{{A^{\kappa}}}
\def\Amptpsigma{{A^{\sigma}}}
\def\Amptpbarkappa{\kern 0.18em\overline{\kern -0.18em A}{}^{\kappa}{}}
\def\Amptpbarsigma{\kern 0.18em\overline{\kern -0.18em A}{}^{\sigma}{}}
\def\AmpAll{|{\cal A}_{3\pi}^\pm(\dmt)|^2}
\def\AmpAllp{|{\cal A}_{3\pi}^+(\dmt)|^2}
\def\AmpAllm{|{\cal A}_{3\pi}^-(\dmt)|^2}

\def\Tbarkappa{\kern 0.18em\overline{\kern -0.18em T}{}^{\kappa}{}}
\def\Tbarsigma{\kern 0.18em\overline{\kern -0.18em T}{}^{\sigma}{}}
\def\Pbarkappa{\kern 0.18em\overline{\kern -0.18em P}{}^{\kappa}{}}
\def\Pbarsigma{\kern 0.18em\overline{\kern -0.18em P}{}^{\sigma}{}}
\def\kappab{\overline\kappa}

\def\ji{\kappab}
\def\ij{\kappa}
\def\Aij{{A^{\ij}}}
\def\Abij{\kern 0.18em\overline{\kern -0.18em A}{}^{\ij}{}}
\def\Tij{{T^{\ij}}}
\def\Tji{{T^{\ji}}}
\def\Pij{{P^{\ij}}}
\def\Pji{{P^{\ji}}}
\def\R{{\rm Re}}
\def\I{{\rm Im}}
\def\kappm{\kappa^{+-}}
\def\kapmp{\kappa^{-+}}

\def\bbar{\Bz \Bzb}
\def\Czz{C_{\rho\pi}^{00}}
\def\Szz{S_{\rho\pi}^{00}}
\def\Crhopi{C}
\def\dCrhopi{\Delta C}

\def\dS{\Delta S}
\def\dC{\Delta C}
\def\rhopi{\rho\pi}
\def\Acp{{\cal A}_{\rho\pi}}

\def\Acppm{{\cal A}_{\rho\pi}^{+-}}
\def\Acpmp{{\cal A}_{\rho\pi}^{-+}}

\def\Nbpm{{\kern 0.18em\overline{\kern -0.18em N}}^{+-}}
\def\Nbmp{{\kern 0.18em\overline{\kern -0.18em N}}^{-+}}
\def\rar{\rightarrow}

\def\U{{\mathfrak U}}
\def\C{{\mathfrak C}}
\def\S{{\mathfrak S}}

\def\Mu{\mu}
\def\Chi2MinaMu{\chi^2_{\min ;\a,\Mu}}
\def\Chi2MinMu{\chi^2_{\min ;\Mu}(a)}

\def\dmt{\Delta t}
\def\dmd{\Delta m_d}

\def\TM{{\rm TM}}
\def\SCF{{\rm SCF}}

\def\fscfave{\kern 0.18em\overline{\kern -0.18em f}_{\rm SCF}}
\def\fscf{f_{\rm SCF}}
\def\fscfi{f_{{\rm SCF},i}}

\def\abar{\bar{a}}

\def\Bbar{\kern 0.18em\overline{\kern -0.18em B}{}\xspace}

\def\BRpmb{{\cal \kern 0.18em\overline{\kern -0.18em  B}}{}_{\rho\pi}^{+-}}
\def\BRmpb{{\cal \kern 0.18em\overline{\kern -0.18em  B}}{}_{\rho\pi}^{-+}}

\def\BRipmb{{\cal \kern 0.18em\overline{\kern -0.18em  B}}{}_{\rho^+\pi^-}}
\def\BRimpb{{\cal \kern 0.18em\overline{\kern -0.18em  B}}{}_{\rho^-\pi^+}}

\def\Abar{\kern 0.18em\overline{\kern -0.18em A}{}}
\def\abar{\kern 0.18em\overline{\kern -0.18em a}{}}

\def\Apm{A^{+}}
\def\Amp{A^{-}}
\def\Apmb{\Abar^{+}}
\def\Ampb{\Abar^{-}}

\def\Azz{A^{0}}
\def\Azzb{\Abar^{0}}

\def\ie{{\em i.e.}} 
\def\cf{{\em cf.}} 
\def\eg{{\em e.g.}}

\newcommand{\UPm    }{\ensuremath{U_+^-}}
\newcommand{\UMp    }{\ensuremath{U_-^+}}
\newcommand{\UMm    }{\ensuremath{U_-^-}}
\newcommand{\UPMpRe }{\ensuremath{U_{+-}^{+,\R}}}
\newcommand{\UPMmRe }{\ensuremath{U_{+-}^{-,\R}}}
\newcommand{\UPMpIm }{\ensuremath{U_{+-}^{+,\I}}}
\newcommand{\UPMmIm }{\ensuremath{U_{+-}^{-,\I}}}
\newcommand{\Uzp    }{\ensuremath{U_0^+}}
\newcommand{\UPzpRe }{\ensuremath{U_{+0}^{+,\R}}}
\newcommand{\UPzpIm }{\ensuremath{U_{+0}^{+,\I}}}
\newcommand{\UMzpRe }{\ensuremath{U_{-0}^{+,\R}}}
\newcommand{\UMzpIm }{\ensuremath{U_{-0}^{+,\I}}}
\newcommand{\IP     }{\ensuremath{I_+}}
\newcommand{\IM     }{\ensuremath{I_-}}
\newcommand{\IPMRe  }{\ensuremath{I_{+-}^{\R}}}
\newcommand{\IPMIm  }{\ensuremath{I_{+-}^{\I}}}

\newcommand{\Iz     }{\ensuremath{I_{0}}}
\newcommand{\IMzIm  }{\ensuremath{I_{-0}^{\I}}}
\newcommand{\IMzRe  }{\ensuremath{I_{-0}^{\R}}}
\newcommand{\IPzIm  }{\ensuremath{I_{+0}^{\I}}}
\newcommand{\IPzRe  }{\ensuremath{I_{+0}^{\R}}}
\newcommand{\Uzm    }{\ensuremath{U_{0}^{-}}}
\newcommand{\UMzmRe }{\ensuremath{U_{-0}^{-,\R}}}
\newcommand{\UMzmIm }{\ensuremath{U_{-0}^{-,\I}}}
\newcommand{\UPzmRe }{\ensuremath{U_{+0}^{-,\R}}}
\newcommand{\UPzmIm }{\ensuremath{U_{+0}^{-,\I}}}

\newcommand{\sigPiNb}{\ensuremath{N_{\rm 3 \pi}}}
\newcommand{\fzz    }{\ensuremath{f(\rho^0\piz)}}

%% file: pubboard/authors_jan2007.tex
%% author list as of 05-Jan-2007 (582 authors)
%
\author{B.~Aubert}
\author{M.~Bona}
\author{D.~Boutigny}
\author{Y.~Karyotakis}
\author{J.~P.~Lees}
\author{V.~Poireau}
\author{X.~Prudent}
\author{V.~Tisserand}
\author{A.~Zghiche}
\affiliation{Laboratoire de Physique des Particules, IN2P3/CNRS et Universit\'e de Savoie, F-74941 Annecy-Le-Vieux, France }
\author{J.~Garra~Tico}
\author{E.~Grauges}
\affiliation{Universitat de Barcelona, Facultat de Fisica, Departament ECM, E-08028 Barcelona, Spain }
% ...
\collaboration{The \babar\ Collaboration}
\noaffiliation

%% file: Introduction.tex
\section{INTRODUCTION}
\label{sec:Introduction}

Measurements of the parameter $\stwob$~\cite{BabarS2b,BelleSin2beta} 
have established \CP violation in the $\Bz$ meson system. 
These measurements provide strong support for the Kobayashi and Maskawa model of 
this phenomenon as arising from a single phase in the three-generation
CKM quark-mixing matrix~\cite{CKM}. 
We present in this paper results from a time-dependent 
analysis of the $\Bz\to\pip\pim\piz$ \cite{ccref} Dalitz plot  which is dominated 
by  intermediate vector resonances ($\rho$). The goal of this analysis
is the simultaneous extraction of the strong transition 
amplitudes and the weak interaction phase 
$\alpha\equiv \arg\left[-V_{td}^{}V_{tb}^{*}/V_{ud}^{}V_{ub}^{*}\right]$
of the Unitarity Triangle. In the Standard Model, a non-zero 
value for $\alpha$ is responsible for the occurrence 
of mixing-induced \CP violation in this decay.
The \babar\  and Belle experiments have obtained constraints
on $\alpha$ from the measurement of effective quantities $\stwoa_{\rm eff}$
in $B$ decays to $\pip\pim$~\cite{babarpipi,bellepipi} and 
from  $\rho^+\rho^-$~\cite{babarrhorho,bellerhorho}, using an 
isospin analysis~\cite{GLisospin}.

Unlike $\pip\pim$, $\rho^{\pm}\pi^{\mp}$ is not a \CP 
eigenstate and four flavor-charge configurations
$(\Bz(\Bzb) \to \rho^{\pm}\pi^{\mp})$ must be considered.  
The corresponding isospin analysis~\cite{Lipkinetal} is unfruitful
with the present data sample since two pentagonal amplitude relations with 
12 unknowns have to be solved (compared to 6 unknowns for
the  $\pip\pim$ and $\rho^+\rho^-$ systems). However, it
has been pointed out by Snyder and Quinn~\cite{SnyderQuinn} that 
one can obtain the necessary degrees of freedom to constrain
$\alpha$ without ambiguity by explicitly including in the analysis the 
variation of the strong phases of the interfering $\rho$ resonances in the Dalitz plot.

\subsection{DECAY AMPLITUDES}
\label{sec:kinmeatics}

We consider the decay of a spin-zero $\Bz$ meson with four-momentum
$p_B$ into the three daughters $\pip$, $\pim$, $\piz$,
with $p_+$, $p_-$, and $p_0$ their corresponding four-momenta. We take
as the independent (Mandelstam) variables the invariant squared masses of the 
charged and neutral pions
\beq
\label{eq:dalitzVariables}
       \spz \;=\; (p_+ + p_0)^2~, \hspace{1cm}
       \smz \;=\; (p_- + p_0)^2~.
\eeq
The invariant squared mass of the positive and negative pion, 
$\spm \;=\; (p_+ + p_-)^2$, is obtained, from energy and 
momentum conservation,
\beq
\label{eq:magicSum}
	\spm \;=\; \mBz^2 + 2m_{\pi^+}^2 + m_{\pi^0}^2
		   - \spz - \smz~.
\eeq
The differential $\Bz$ decay rate distribution as a function of the 
variables defined in Eq.~(\ref{eq:dalitzVariables}) (\ie, the 
{\em Dalitz plot}) reads
\beq
\label{eq:partialWidth}
	d\Gamma(\Btopipipi) \;=\; 
	\frac{1}{(2\pi)^3}\frac{|\Amptp|^2}{32 \mBz^3}\,d\spz d\smz~,
\eeq
where $\Amptp$ is the Lorentz-invariant amplitude
of the three-body decay \cite{PDG}. 

We assume in the following that the amplitude $\Amptp$ and its $CP$ 
conjugate $\Amptpbar$, corresponding to the transitions $\Bz\to\pip\pim\piz$ 
and $\Bzb\to\pip\pim\piz$, respectively, are dominated by the three 
resonances $\rho^+$, $\rho^-$ and $\rho^0$. The $\rho$ resonances
are assumed to be the sum of the ground state $\rho(770)$ and the
radial excitations $\rho(1450)$ and $\rho(1700)$, with 
masses and widths determined by a combined fit 
to $\tau^+\to\nutb\pip\piz$ and $\epem\to\pip\pim$ data~\cite{taueeref}.
Since the hadronic environment is different in \B decays, we 
do not rely on this result for the relative $\rho(1450)$
and $\rho(1700)$
amplitudes but instead simultaneously measure them with the \CP parameters from the fit. Variations of
the other parameters and possible contributions to the $\Bz\to\pip\pim\piz$ 
decay other than the $\rho$ resonances are studied as part of the systematic 
uncertainties (Section~\ref{sec:Systematics}).

We write the  $\Amptp$ and $\Amptpbar$ amplitudes  \cite{SnyderQuinn,BaBarPhysBook}
\beqn
\label{eq:amp}
   \Amptp    		
	&=& \fpz \Apm + \fmz \Amp + \fpm\Azz ~, \\
\label{eq:ampBar}
   \Amptpbar 	
	&=& \fpz \Apmb + \fmz \Ampb + \fpm\Azzb ~,
\eeqn
where the $f_\kappa$ (with $\kappa=\{+,-,0\}$ denoting the charge of the
$\rho$ from the decay of the $\Bz$ meson) 
are functions of the Dalitz variables 
$\spz$ and $\smz$ that incorporate the kinematic and dynamical properties 
of the $\Bz$ decay into a vector $\rho$ resonance and a 
pseudoscalar pion. The $\Aij$ are 
complex amplitudes
that include weak and strong transition phases and that are independent 
of the Dalitz variables. 

Following Ref.~\cite{taueeref}, the $\rho$
resonances are  parameterized in  $f_\kappa$ (where $\kappa$ is the charge) as a sum of the $\rho(770)$, $\rho(1450)$, 
and  $\rho(1700)$ resonances:  
\beqn
\label{eq:eetauGSff}
        f_\kappa(s) \;\propto\; &
                    F_{\rho(770)}(s)
                        + \ampfrac_{\rho^\prime}
                          e^{i\phi_{\rho^\prime}}
                          F_{\rho(1450)}(s) 
                        + \\\nonumber
	&	\ampfrac_{\rho^{\prime\prime}}
                          e^{i\phi_{\rho^{\prime\prime}}}
                          F_{\rho(1700)}(s)~,
\eeqn
where the $F_{\rho}$ are modified relativistic Breit-Wigner 
functions introduced by Gounaris and Sakurai (GS)~\cite{rhoGS} 
and the $a_{\rho}$ ($\phi_{\rho}$) are the  magnitudes
(phases) of the higher mass $\rho$ resonances, relative to the $\rho(770)$. In 
this analysis, we assume that the $a$ and $\phi$ for $f_+$ and $f_-$ are the 
same while for $f_0$, corresponding to the much smaller $\rho^0$ component, we fix 
$a_\rho^\prime$ and  $a_\rho^{\prime\prime}$ to zero.
Note that the definitions~(\ref{eq:amp})
and (\ref{eq:ampBar}) are based on the assumption that the relative phases 
between the $\rho(770)$ and its radial excitations are \CP-conserving.

Due to angular 
momentum conservation, the spin-one $\rho$ resonance is restricted to 
a helicity-zero state. For a $\rho^{\a}$ resonance with charge $\a$,
the GS function is multiplied by the kinematic function 
$-4|{\bf p}_\a||{\bf p}_\c|\cos\theta_{\a}$, where ${ \bf p}_\a$ 
is the momentum of either of the daughters of the $\rho$ resonance, 
 ${\bf p}_\c$ is the momentum of the 
particle not from the $\rho$ decay,
and $\cos\theta_{\a}$ is the cosine of the helicity angle of
the $\rho^{\a}$ all defined, in the $\rho$-resonance rest frame. 
For the $\rho^+$ ($\rho^-$), $\theta_{+}$ 
($\theta_{-}$) is defined by the angle between the $\pi^0$ ($\pi^-$) 
 momentum in the $\rho^+$ ($\rho^-$) rest frame and the $\rho^+$ ($\rho^-$) 
flight direction in the $\Bz$ rest frame. For the $\rho^0$, 
$\theta_{0}$ is defined by the angle between the $\pi^+$ momentum in 
the $\rho^0$ rest frame and the $\rho^0$ flight direction in 
the $\Bz$ rest frame. With these definitions, each pair of GS functions 
interferes destructively at equal masses.

The factor of $\cos\theta_{\a}$ 
in the kinematic functions leads to an increased population in the interference 
regions between the different $\rho$ bands in the Dalitz plot, and thus 
increases the sensitivity of this analysis~\cite{SnyderQuinn}.

\subsection{TIME DEPENDENCE}

With $\deltat \equiv t_{\tpi} - t_{\rm tag}$ defined as the proper 
time interval between the decay of the fully reconstructed $B^0_{\tpi}$ 
and that of the  other meson $\Bz_{\rm tag}$ from the \FourS,  the time-dependent decay
rate $\AmpAllp$ ($\AmpAllm$) when the $\Bz_{\rm tag}$ is a $\Bz$ ($\Bzb$) 
is given by 
\beqn
\label{eq:dt}
    \AmpAll&
	=&
		\frac{e^{-|\dmt|/\tau_{B^0}}}{4\tau_{B^0}}
	\bigg[\absAmptp^2 + \absAmptpbar^2\nonumber\\
&&	      \mp \left(\absAmptp^2 - \absAmptpbar^2\right)\cos(\dmd\dmt)\nonumber\\
&&	      \pm\,2\I\left[\frac{q}{p}\Amptpbar\Amptp^*\right]\sin(\dmd\dmt)	
	\bigg]~,
\eeqn
where $\tau_{B^0}$ is the mean neutral $B$ lifetime and $\deltamd$ is the $\BzBzb$ 
mass difference. Here, we have assumed that \CP violation in $\bbar$ 
mixing is absent ($|q/p|=1$) and the lifetime difference between $B_H$
and $B_L$ is $\Delta\Gamma_{B_d}=0$.
Inserting the amplitudes~(\ref{eq:amp}) and (\ref{eq:ampBar}), one 
obtains for the terms in Eq.~(\ref{eq:dt})
\beqn
   \label{eq:UI}
   \absAmptp^2 \pm \absAmptpbar^2 = \sum_{\kappa\in\{+,-,0\}}  |f_\kappa|^2U_\kappa^\pm + \nonumber\\
	\bk\bk\sum_{\kappa <\sigma\in\{+,-,0\}} 
	2\left(
	    \,\R\left[f_\kappa f_\sigma^*\right]U_{\kappa\sigma}^{\pm,\R}
	 \ - \,\I\left[f_\kappa f_\sigma^*\right]U_{\kappa\sigma}^{\pm,\I}
	\right)~,
	\nonumber\\[0.2cm]
   \I\left(\frac{q}{p}\Amptpbar\Amptp^*\right)
	=
	\sum_{\kappa\in\{+,-,0\}}  |f_\kappa|^2I_\kappa \;\;+ \nonumber\\ 
	\sum_{\kappa <\sigma\in\{+,-,0\}} 
	\left(
	    \,\R\left[f_\kappa f_\sigma^*\right]I_{\kappa\sigma}^{\I}
	  + \,\I\left[f_\kappa f_\sigma^*\right]I_{\kappa\sigma}^{\R}
	\right)~,	
\eeqn
with
\beqn
\label{eq:firstObs}
   U_\kappa^\pm 		&=& |\Amptpkappa|^2 \pm |\Amptpbarkappa|^2~, \\
   U_{\kappa\sigma}^{\pm,\R}&=& \R\left[\Amptpkappa \Amptpsigma{}^* 
				    \pm \Amptpbarkappa \Amptpbarsigma{}^*\right]~, \\
   U_{\kappa\sigma}^{\pm,\I}&=& \I\left[\Amptpkappa \Amptpsigma{}^* 
				    \pm \Amptpbarkappa \Amptpbarsigma{}^*\right]~, \\
   I_\kappa			&=& \I\left[\Amptpbarkappa\Amptpkappa{}^*\right]~, \\
   I_{\kappa\sigma}^{\R}  	&=& \R\left[\Amptpbarkappa\Amptpsigma{}^* 
					    - \Amptpbarsigma\Amptpkappa{}^*\right]~, \\
\label{eq:lastObs}
   I_{\kappa\sigma}^{\I}  	&=& \I\left[\Amptpbarkappa\Amptpsigma{}^* 
					    + \Amptpbarsigma\Amptpkappa{}^*\right]~.
\eeqn

The 27 coefficients~(\ref{eq:firstObs})--(\ref{eq:lastObs}) are real-valued
parameters that multiply the $f_\kappa f_\sigma^*$ bilinears (where $\kappa$
and $\sigma$ denote the charge of the $\rho$ resonances)~\cite{quinnsilva}. 
These coefficients are the 
observables that are determined by the fit. Each of the coefficients 
is related in a unique way to physically more intuitive quantities,
such as tree-level and penguin-type amplitudes, the angle $\alpha$, or
the quasi-two-body \CP and dilution parameters~\cite{rhopipaper} 
(\cf\   Section~\ref{sec:Physics}). The parameterization~(\ref{eq:UI}) is 
general: the information on the mirror solutions (\eg, on the angle $\alpha$)
that are present in the transition amplitudes $\Aij$, $\Abij$ is conserved.

The decay rate~(\ref{eq:dt}) is used as a probability density 
function (PDF) in a maximum-likelihood fit and must therefore be normalized:
\beq
 	\AmpAll \;\longrightarrow\;
	\frac{1}{\langle |\Amptp|^2 + |\Amptpbar|^2 \rangle }\AmpAll~,
\eeq
where
\beqn
\label{eq:Norm}
	\langle |\Amptp|^2 + |\Amptpbar|^2 \rangle
	\;=\;
	\sum_{\kappa\in\{+,-,0\}}  \langle|f_\kappa|^2\rangle U_\kappa^+
	\;\nonumber\\
	+\;
	2\R\!\!\!\!\!\!\!\!
		\sum_{\kappa <\sigma\in\{+,-,0\}} \!\!\!\!
			\langle f_\kappa f_\sigma^*\rangle
			\left(
				U^{+,\mathrm{Re}}_{\kappa\sigma} +
				i\cdot U^{+,\mathrm{Im}}_{\kappa\sigma}
			\right)	   
	~,
\eeqn
where $\langle ... \rangle$ denotes the expectation value over the Dalitz plot. 
The complex expectation values $\langle f_\kappa f_\sigma^*\rangle$ are 
obtained from  Monte Carlo integration of the Dalitz 
plot~(\ref{eq:partialWidth}), taking into account acceptance and 
resolution effects.
In this paper, we determine the relative values of $U$ and $I$ coefficients 
to $U_+^+$ leaving 26 free coefficients.

The choice to fit for the $U$ and $I$ coefficients rather than
fitting for the complex transition amplitudes and the weak phase $\alpha$
directly is motivated by the following technical
simplifications: $(i)$ in contrast to the amplitudes, there is a unique
solution for the $U$ and $I$ coefficients requiring only a single fit
to the selected data sample;
$(ii)$ in the presence of background, we find that the errors on the $U$ and $I$ coefficients are
approximately Gaussian, which in general is not the case
for the amplitudes; and $(iii)$ the propagation of systematic uncertainties
and the averaging between different measurements are straightforward for
the $U$ and $I$ coefficients.

The $U_\kappa^+$ coefficients are related to resonance branching
fractions and charge asymmetries; the $U_\kappa^-$ coefficients 
determine the relative 
abundance of the $\Bz$ decay into $\rho^+\pim$ and $\rho^-\pip$ %(dilution)
and the time-dependent direct \CP asymmetries. The $I_\kappa$ measure 
mixing-induced \CP violation and are sensitive to strong phase shifts.
Finally, the $U_{\kappa\sigma}^{\pm,\R(\I)}$ and $I_{\kappa\sigma}^{\R(\I)}$
coefficients describe the interference pattern in the Dalitz plot, and their presence
distinguishes this analysis from the quasi-two-body analysis previously
reported in Ref.~\cite{rhopipaper}.
They represent the additional degrees of freedom that allow one to 
determine the unknown penguin contribution and the relative strong phases.
However, because 
the overlap regions of the resonances are small and because the events
reconstructed in these regions suffer from large misreconstruction 
rates and background, a substantial data sample is needed to perform a 
fit that constrains all amplitude parameters.

We determine the physically relevant  quantities in a subsequent least-squares
fit to the measured $U$ and $I$ coefficients.

\subsection{THE SQUARE DALITZ PLOT}
\label{sec:SquareDP}

\begin{figure}[b]
  \centerline{ \epsfxsize8.2cm\epsffile{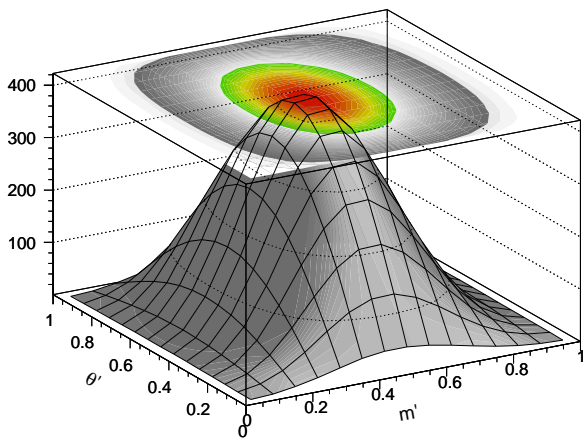}}
  \caption{\label{fig:jacobian}
	Jacobian determinant~(\ref{eq:detJ}) of the 
	transformation~(\ref{eq:SqDalitzTrans}) defining the square Dalitz plot (SDP).
	Such a distribution would be obtained in the SDP if events were uniformly
	distributed over the nominal Dalitz plot.}
\end{figure}

\begin{figure*}[t]
  \centerline{\epsfxsize8.0cm\epsffile{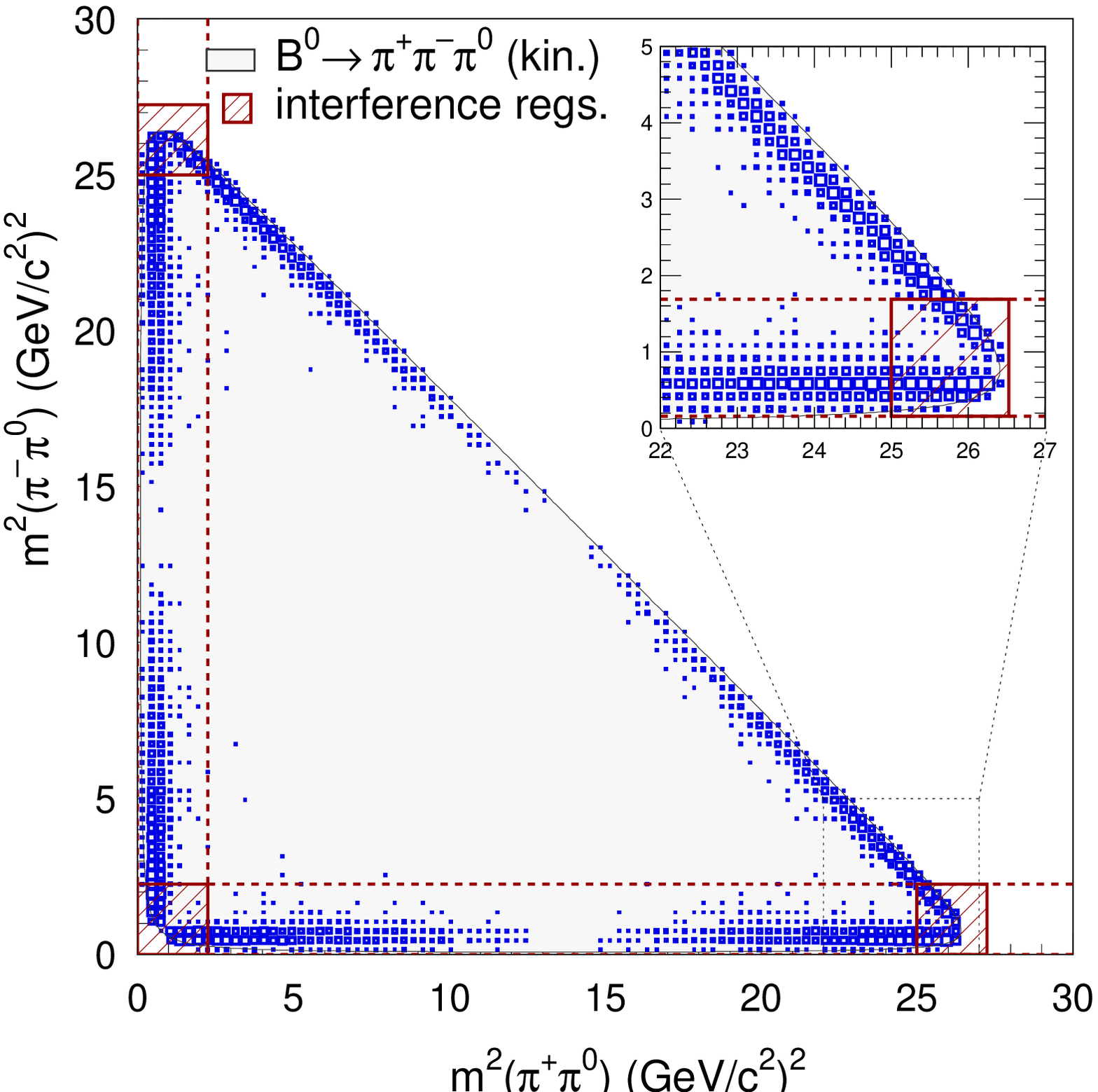}
	      \epsfxsize8.0cm\epsffile{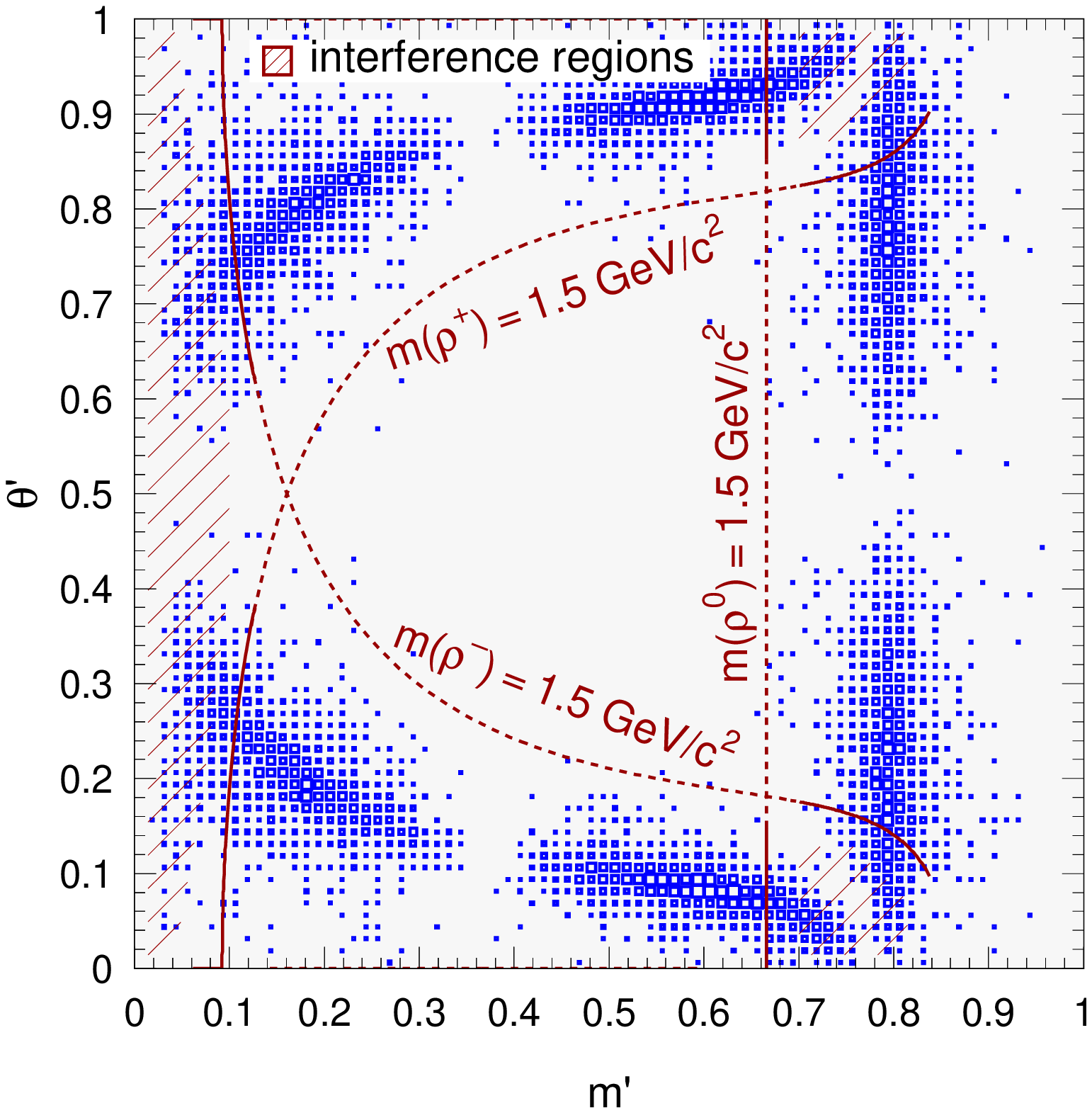}}
 \caption{\label{fig:DPs}
	Nominal (left) and square (right) Dalitz plots for Monte-Carlo
	generated $\Bz\rar\pi^+\pi^-\pi^0$ decays. Comparing the two
        Dalitz plots shows that the transformation~(\ref{eq:SqDalitzTrans})
        indeed homogenizes the distribution of events, which are no longer
  	near the plot boundaries but rather cover a larger fraction of
	the physical region. The decays have been
	simulated without any detector effect and the amplitudes $\Apm$,
	$\Amp$ and $\Azz$ have all been chosen equal to 1 in order to
	have destructive interferences where the  $\rho$ bands overlap. The main
	overlap regions between the  $\rho$ bands
	are indicated by the hatched areas.
	Dashed lines in both plots correspond to
	$\sqrt{s_{+,-,0}}=1.5~{\rm GeV}/c^2$; the central region of the Dalitz
	plot (defined by requiring that all 3 two-body invariant masses exceed
	this threshold) contains few signal events.}
\end{figure*}

Both the signal events and the combinatorial $\epem\to q\bar q$ ($q=u,d,s,c$) 
continuum background events populate the kinematic boundaries of the 
Dalitz plot due to the low final state masses compared with the $\Bz$ mass. 
We find the representation~(\ref{eq:partialWidth}) is inconvenient when one wants 
to use empirical reference shapes in a maximum-likelihood fit.
Large variations occurring in small areas of the Dalitz plot are very difficult to describe in detail.These regions are particularly important since it is here that 
the interference between $\rho$-resonances, 
and hence our ability to determine the strong phases, occurs. 
We therefore apply the transformation
\beq
\label{eq:SqDalitzTrans}
	d\spz \,d\smz \;\longrightarrow \detJ\, d\mprime\, d\thetaprime~,
\eeq
which defines the {\em Square Dalitz plot} (SDP). The new coordinates 
are
\beq
\label{eq:SqDalitzVars}
	\mprime \equiv \frac{1}{\pi}
		\arccos\left(2\frac{\mpm - \mpmMin}{\mpmMax - \mpmMin}
			- 1
		      \right),~
	\thetaprime \equiv \frac{1}{\pi}\theta_{0},
\eeq
where $\mpm=\sqrt{s_0}$ is the invariant mass of the charged particles,
$\mpmMax=\mBz - m_{\pi^0}$ and $\mpmMin=2m_{\pi^+}$ are the kinematic
limits of $\mpm$, $\theta_{0}$ is the $\rho^0$ helicity angle,
and $J$ is the Jacobian of the transformation.
Both new variables  range between 0 and 1.
The determinant of the Jacobian is given by
\beq
\label{eq:detJ}
	\detJ \;=\;	4 \,|{\bf p}^*_+||{\bf p}^*_0| \,\mpm
			\cdot 	
			\frac{\partial \mpm}{\partial \mprime}
			\cdot 	
			\frac{\partial \cos\theta_{0}}{\partial \thetaprime}~,
\eeq
where 
$|{\bf p}^*_+|=\sqrt{E^*_+ - m_{\pi^+}^2}$ and
$|{\bf p}^*_0|=\sqrt{E^*_0 - m_{\pi^0}^2}$, and where the energies 
$E^*_+$ and $E^*_0$ are defined in the $\pi^+\pi^-$ rest frame.
Figure~\ref{fig:jacobian} shows the determinant of the Jacobian as a function
of the SDP parameters $\mprime$ and $\thetaprime$. If the events in the
nominal Dalitz plot were distributed according to a uniform (non-resonant)
three-body phases space, their distribution in the SDP would match the plot of $\detJ$.

The effect of the transformation~(\ref{eq:SqDalitzTrans}) is illustrated in  Fig.~\ref{fig:DPs}, which displays the nominal and square Dalitz
plots for simulated signal events generated with Monte Carlo.
As is shown, the transformation is benificial because:
($i$) it expands the regions of interference so that equal size
bins cover this region in more detail; and ($ii$) it avoids the
curved edge of bins on the boundary.
This simulation does not take into
account any detector effects and corresponds to a particular choice of the
decay amplitudes for which destructive interferences occur where the $\rho$
resonances overlap. To simplify the comparison, hatched areas showing the 
interference regions between $\rho$ bands and dashed isocontours
$\sqrt{s_{+,-,0}}=1.5~{\rm GeV}/c^2$ have been superimposed on both Dalitz
plots.

%% file: DetectorAndData.tex
\section{THE \babar\ DETECTOR AND DATASET}
\label{sec:babar}

The data used in this analysis were collected with the \babar\ 
detector at the \pep2\ asymmetric-energy $e^+e^-$ storage ring at 
SLAC between October 1999 and August 2006. The sample consists of about
$346\;\mathrm{fb}^{-1}$, corresponding to $(375\pm4)\times10^{6}$ 
$B\Bbar$ pairs collected at the \FourS resonance (``on-resonance''), 
and an integrated luminosity of 21.6~\invfb collected about 40~\mev 
below the~\FourS (``off-resonance'').  In addition, we use GEANT4 \cite{geant} 
simulated Monte Carlo (MC) events to study detector efficiency and 
backgrounds. 

%%% put something on the signal monte carlo

A detailed description of the \babar\ detector is presented in 
Ref.~\cite{babarNim}. The tracking system used for charged particle and vertex 
reconstruction has two main components: a silicon vertex tracker 
(SVT) and a drift chamber (DCH), both operating within a 1.5-T 
magnetic field generated by a superconducting solenoid. 
Photons are identified in an electromagnetic calorimeter (EMC) 
surrounding a detector of internally reflected Cherenkov light 
(DIRC), which associates Cherenkov photons with tracks for particle 
identification (PID). Muon candidates are identified with the
use of the instrumented flux return (IFR) of the solenoid.

%% file: AnalysisMethod.tex
\section{ANALYSIS METHOD}
\label{sec:Analysis}

The $U$ and $I$ coefficients and the $\Btopipipi$ event yield are
determined by a maximum-likelihood fit of the signal and background model to the 
selected candidate events. Kinematic and event shape variables 
exploiting the characteristic properties of the events are used 
in the fit to discriminate signal from background. 

\subsection{EVENT SELECTION AND BACKGROUND SUPPRESSION}
\label{subsec:selection}

We reconstruct $\Btopipipi$ candidates from pairs of 
oppositely-charged tracks 
and a $\pi^0\to\gamma\gamma$ candidate.  In order to ensure that all events are within 
the Dalitz plot boundary, we constrain the three-pion invariant mass to the $B$ mass after final selections have been made.  
The largest source of background is from continuum $\epem\to q\overline{q}$ production.  
We use information from the tracking system, EMC, and DIRC to 
remove tracks for which the PID is consistent with the electron, kaon, 
or proton hypotheses. In addition, we require that at least one track 
has a signature in the IFR that is inconsistent with the muon 
hypothesis.  This selection retains 92\% of signal events while rejecting 
42\% of continuum background events.  
The $\pi^0$ candidate mass $m(\gamma\gamma)$ must satisfy $0.11<m(\gamma\gamma)<0.16\gevcc$, 
where each photon, $\gamma$,  is required to have an energy greater than $50\mev$
in the laboratory frame (LAB) and to exhibit a lateral profile of energy 
deposition in the EMC consistent with an electromagnetic shower.

A $B$-meson candidate is characterized kinematically by the beam-energy substituted 
mass 
$\mes=\sqrt{{(E^{\rm cm}_{\rm beam})^2}-(p_B^{\rm cm})^2}$ 
and energy difference $\de = E_B^*-\half\sqrt{s}$, 
where $(E_B,\pvec_B)$ and $(E_0,\pvec_0)$ are the four-vectors
of the $B$-candidate and the initial electron-positron systems
respectively. The asterisk denotes the center-of-mass (CM) frame
and $s$ is the square of the CM energy.
We require $5.272 < \mes <5.288\gevcc$, which retains $81\%$
of the signal and $8\%$ of the continuum background events. 
The $\de$ resolution 
exhibits a dependence on the $\pi^0$ energy and therefore varies 
across the Dalitz plot. To avoid bias in the Dalitz plot, we introduce
the transformed quantity $\deprime=(2\de - \demax - \demin)/(\demax - \demin)$,
with $\deminmax(\mpm)=c_{\pm}-\left(c_{\pm}\mp\bar c\right)(\mpm/\mpmMax)^2$,
where $\mpm=\sqrt{s_0}$ is strongly correlated with the energy of the $\piz$. 
We use the values
$\bar c = 0.045\gev$, $c_{-} = -0.140\gev$, $c_{+} = 0.080\gev$,
$\mpmMax = 5.0\gev$, and require $-1<\deprime<1$. 
These values have been obtained from Monte Carlo simulation.
The requirement retains $75\%$ ($25\%$) of the signal (continuum) events.

Backgrounds arise primarily from random combinations of $\pi^\pm$ and $\pi^0$ 
candidates in continuum events.
Continuum events tend to  have a more ``jet-like'' structure than  
$B$ decays which are produced nearly at rest in the CM system.  
To enhance discrimination between signal and continuum, we 
use a neural network (NN)~\cite{NNo} to combine four discriminating variables: 
the angles with respect to the beam axis of the $B$ momentum and $B$ thrust 
axis in the \FourS\ frame, and the zeroth and second order polynomials
$L_{0,2}$ of the energy flow about the $B$ thrust axis.  The polynomials
are defined by $ L_n = \sum_i {p}_i\cdot\left|\cos\theta_i\right|^n$,
where $\theta_i$ is the angle with respect to the $B$ thrust axis of any
track or neutral cluster $i$, ${\bf p}_i$ is its momentum, and the sum
excludes the $B$ candidate.  
The NN is trained with off-peak data and
simulated signal events. The final sample of signal candidates 
is selected with a requirement on the NN output that retains $77\%$ ($8\%$) 
of the signal (continuum) events.  A total of 35444  on-peak data events pass the 
selection.  

The time difference $\deltat$ is obtained from the measured distance between 
the $z$ positions (along the beam direction) of the $\Bz_{\tpi}$ and 
$\Bz_{\rm tag}$ decay vertices, and the boost $\beta\gamma=0.56$ of 
the \epem\ system: $\deltat = \Delta z/\beta\gamma c$.  The $\Bz_{\rm tag}$
vertex is determined from the charged particles in the event not included 
in the  signal $B$. 
To determine the flavor of the $\Bz_{\rm tag}$ 
we use the $B$ flavor-tagging algorithm of Ref.~\cite{BabarS2b}.
This produces six mutually exclusive tagging categories. We improve the efficiency 
of the signal selection by retaining untagged events in a seventh category
which  contribute to the measurement of direct \CP violation. 

Multiple \B candidates passing the full selection occur 
in $16\%$ $(\rho^\pm\pi^\mp)$ and $9\%$ $(\rho^0\pi^0)$ 
 of $\rho(770)$ MC events. 
If the multiple candidates have different $\pi^0$ candidates, 
we choose the \B candidate with the reconstructed $\pi^0$ mass closest 
to the nominal $\pi^0$ mass; 
in the case that more than one candidate have the same $\pi^0$, we 
arbitrarily chose a  reconstructed 
\B candidates passing the selection (this occurs in $4\%$ of events).  

\begin{figure}[tbh]
  \centerline{ \epsfxsize8.2cm\epsffile{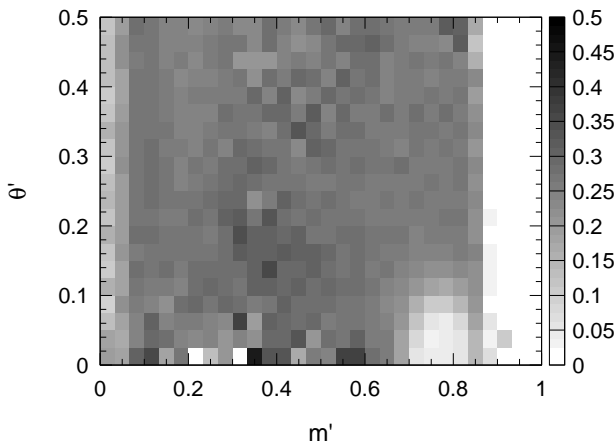}}
  \caption{\label{fig:sigeff}
        The signal efficiency distribution on the square Dalitz plot.  Note that
	the plot is folded in $\thetaprime$ since the distribution 
	is nearly symmetric in this variable. }
\end{figure}

The signal efficiency determined from MC simulation is $24\%$ for 
$B^0 \to \rho^\pm\pi^\mp$ and $B^0 \to \rho^0\pi^0$ events, and 
$11\%$ for non-resonant $\Btopipipi$ events. The signal efficiency distribution on the SPD is shown in Figure \ref{fig:sigeff}.

 The signal events passing the event selection are a combination of correctly 
reconstructed (``truth-matched'', TM) events and  mis-reconstructed 
(``self-cross-feed'', SCF) events.
Of the selected signal events, $22\%$ of $B^0 \to \rho^\pm\pi^\mp$, 
$13\% $ of $B^0 \to \rho^0\pi^0$, and $6\%$ of non-resonant events are 
mis-reconstructed, according to MC.  Mis-reconstructed events occur when a track or 
neutral cluster from the tagging $B$ is assigned to the reconstructed signal candidate. 
This occurs most often for  low-momentum particles and photons; hence the mis-reconstructed
 events are concentrated in the corners of the standard Dalitz plot.  Since these are also the 
areas where the $\rho$ resonances overlap strongly, it is important to model 
the mis-reconstructed events correctly.  The details of the model for 
the distributions of mis-reconstructed 
events  in the Dalitz plot are described in Section \ref{sec:deltaT}.

\subsection{BACKGROUND FROM OTHER {\em B} DECAYS}

\begin{table*}[t]
\begin{center}
\caption{ \label{tab:bbackground}
	Summary of the \B-background modes taken into account for the
	likelihood model. They have been grouped in 20 classes:
 	 charmless $B^+$ (six), charmless $B^0$ (eight), 
	exclusive  charmed $B^0$ (four) and inclusive $B^0$ and
	 charmed $B^+$ decays. Modes with at least two events expected
	after final selection have been included.}
\input{bBackground.tex}
\vspace{-0.2cm}
\end{center}
\end{table*}

We use MC simulated events to study the background from other $B$ 
decays. More than one-hundred channels were considered in 
these studies, of which 29 are  included
in the final likelihood model.  These exclusive \B-background modes are grouped into eighteen 
different classes according to their kinematic and topological
properties: six for  charmless $\B^+$ decays, eight for charmless $B^0$ decays 
and four for exclusive charmed $\B^0$ decays.
Two additional classes account for inclusive $B^0$ and $B^+$
charmed decays.

Table \ref{tab:bbackground} summarizes the twenty background classes that are
used in the fit. For each mode, the expected number of selected events is
computed by multiplying the selection efficiency (estimated using MC
simulated decays) by the branching fraction, scaled to the dataset 
luminosity ($346\;\mathrm{fb}^{-1}$). The world average branching ratios have been
used for the experimentally known decay modes\cite{PDG,HFAG}. When only upper limits are
given, they have been translated into branching ratios including additional conservative hypotheses 
(e.g., 100\% longitudinal polarization for $B\to\rho\rho$ decay) if needed.

\subsection{THE MAXIMUM-LIKELIHOOD FIT}
\label{subsec:ML}

We perform an unbinned extended maximum-likelihood fit to extract
the total $\Btopipipi$ event yield, and the $U$ and $I$ coefficients
defined in Eqs.~(\ref{eq:firstObs})--(\ref{eq:lastObs}). 
The fit uses the variables  $\dt$, $\mprime$, $\thetaprime$, $\mes$, $\deprime$, and NN output 
to discriminate signal from background. The 
$\dt$ distribution is sensitive to mixing-induced \CP violation
but also provides additional continuum-background rejection. 

The selected on-resonance data sample is assumed to consist of signal, 
continuum-background, and \B-background components, separated by the 
flavor and tagging category of the tag side \B decay. 
The probability density function  ${\cal P}_i^\cat$ for 
event $i$ in tagging category $\cat$ is the sum of the probability densities 
of all components, namely
\beqn
\label{eq:theLikelihood}
	{\cal P}_i^\cat
	&\equiv& 
		N_{\tpi} f^\cat_{\tpi}
		\left[ 	(1-\fscfave^\cat){\cal P}_{\tpi-\TM,i}^\cat +
			\fscfave^\cat{\cal P}_{\tpi-\SCF,i}^\cat 
		\right] 
		\nonumber\\[0.3cm]
	&&
		+\; N^\cat_{q\bar q}\frac{1}{2}
		\left(1 + \Qtagi\Atagqq\right){\cal P}_{q\bar q,i}^\cat
		\nonumber \\[0.3cm]
	&&
		+\; \sum_{j=1}^{N^{B^+}_{\rm class}}
		N_{B^+j} f^\cat_{B^+j}
		\frac{1}{2}\left(1 + \Qtagi \Atagj\right){\cal P}_{B^+,ij}^\cat
		\nonumber \\[0.3cm]
	&&
		+\; \sum_{j=1}^{N^{B^0}_{\rm class}}
		N_{B^0j} f^\cat_{B^0j}
		{\cal P}_{B^0,ij}^\cat~,
\eeqn
where
	$N_{\tpi}$ is the total number of $\pi^+\pi^-\pi^0$ signal events 
	in the data sample;
 	$f^\cat_{\tpi}$ is the fraction of signal events that are 
       	 in tagging category $\cat$;
	$\fscfave^\cat$ is the fraction of SCF events in tagging category $\cat$, 
	averaged over the Dalitz plot;
	${\cal P}_{\tpi-\TM,i}^\cat$ and ${\cal P}_{\tpi-\SCF,i}^\cat$
	are the products of PDFs of the discriminating variables used
	in tagging category $\cat$ for TM and SCF
	events, respectively; 
 	$N^\cat_{q\bar q}$ is the number of continuum events that are 
	in tagging category $\cat$;
	$\Qtagi$ is the tag flavor of the event, defined to be 
	$+1$ for a $\Bz_{\rm tag}$ and $-1$ for a $\Bzb_{\rm tag}$; 
	$\Atagqq$ parameterizes possible flavor tag asymmetry in continuum events; 
	${\cal P}_{q\bar q,i}^\cat$ is the continuum PDF for tagging 
	category $\cat$;
	$N^{B^+}_{\rm class}$ ($N^{B^0}_{\rm class}$) is the number of 
	charged (neutral) $B$-related background classes considered in the fit;
	$N_{B^+j}$ ($N_{B^0j}$) is the number of expected events in
	the charged (neutral) $B$-background class $j$;
	$f^\cat_{B^+j}$ ($f^\cat_{B^0j}$) is the fraction of 
	charged (neutral) $B$-background events of class $j$
	that are in tagging category $\cat$;
	$\Atagj$ describes a possible flavor tag asymmetry in the $B^+$ background
	class $j$; 
	${\cal P}_{B^+,ij}^\cat$ is the $B^+$-background PDF for tagging 
	category $\cat$ and class $j$;
	and ${\cal P}_{B^0,ij}^\cat$ is the neutral-$B$-background 
	PDF for tagging category $\cat$ and class $j$.
Correlations between the flavor tag and the position in the Dalitz plot 
 are absorbed in tag-flavor-dependent 
	Dalitz plot PDFs that are used for $B^+$ and continuum
	background.

The PDFs ${\cal P}_{X}^{\cat}$ ($X=\{{\rm TM, SCF}, {q\overline{q}}, {B^+/B^0}\}$)
are the product of the four PDFs of the discriminating variables,
$x_1 = m_{ES}$, $x_2 = \deprime$, and $x_3 = {\rm NN~output}$, and the triplet
$x_4 = \{\mprime, \thetaprime, \deltat\}$:
\beq
\label{eq:likVars}
	{\cal P}_{X,i(j)}^{\cat} \;\equiv\; 
	\prod_{k=1}^4 P_{X,i(j)}^\cat(x_k)~,
\eeq
where $i$ is the event index and $j$ is a $B$-background class.  The extended likelihood over all tagging categories is given by
\beq
	{\cal L} \;\equiv\;  
	\prod_{\cat=1}^{7} e^{-\overline N^\cat}\,
	\prod_{i=1}^{N^\cat} {\cal P}_{i}^\cat~,
\eeq
where $\overline N^\cat$ is the total number of events expected in category 
$\cat$. 

A total of 68 parameters, including the inclusive signal yield $N_{\tpi}$ and  the 26
$U$ and $I$ coefficients from Eq.~(\ref{eq:dt}), are varied in the fit. Most of the
parameters describing the continuum distributions are also free
in the fit.  The parameterizations of the PDFs are described below and are summarized in Tab. \ref{tab:pdfparameterization}.

\begin{table*}[t]
\begin{center}
\caption{ \label{tab:pdfparameterization}
        Summary of PDF parameterizations where G=Gaussian, PX=X-order polynomial, NP=non-parametric, and biCB=bifurcated Crystal Ball. See Section \ref{sec:deltaT} for a detailed description of the Dalitz plot parameterization for signal.}
\input{pdfTable.tex}

\vspace{-0.2cm}
\end{center}
\end{table*}

\subsubsection{\boldmath THE $\dt$ AND DALITZ PLOT PDFS}
\label{sec:deltaT}

	The Dalitz plot PDFs require as input the Dalitz plot-dependent 
	relative selection efficiency $\e=\e(\mprime,\thetaprime)$, 
	and the SCF fraction, $\fscf=\fscf(\mprime,\thetaprime)$.
	Both quantities are taken from MC simulation. 
	Away from the Dalitz plot corners the efficiency is uniform, while it 
	decreases when approaching the corners where one  of the 
	three particles in the final state is almost at rest in the LAB frame so that the 
	acceptance requirements on the particle reconstruction become 
        restrictive.
	Combinatorial backgrounds, and hence SCF fractions, are large in
	the corners
	of the Dalitz plot due to the presence of soft neutral clusters 
	and tracks. 

	For an event~$i$, we define the time-dependent Dalitz plot PDFs
	\beqn
		P_{\tpi-\TM,i}^{c} &\equiv&
		\varepsilon_i\,(1 - \fscfi^{c})\,\detJi\,\AmpAll~,
		\\[0.3cm]
		P_{\tpi-\SCF,\,i}^{c} &\equiv&
		\varepsilon_i\,\fscfi^{c}\,\detJi\,\AmpAll~,
	\eeqn	
	where $P_{\tpi-\TM,i}$ and $P_{\tpi-\SCF,\,i}$ are normalized. The 
	 normalization involves the expectation values 	
	$\langle \varepsilon\,(1-\fscf)\,\detJ \,f^\kappa f^{\sigma*}\rangle$
	and 
	$\langle \varepsilon\,\fscf\,\detJ\, f^\kappa f^{\sigma*}\rangle$
	for TM and SCF events, where the indices $\kappa$, $\sigma$ 
	run over all resonances belonging to the signal model.
	The expectation values are model-dependent and are 
	computed with the use of MC integration over the square Dalitz plot:
	\beqn
	\label{eq:normAverage}
		\langle \varepsilon\,(1-\fscf)\,\detJ\, f^\kappa f^{\sigma*}\rangle
		\;=\; \nonumber\\
		\frac{\int_0^1\int_0^1 
			    \varepsilon\,(1-\fscf)\,\detJ\, f^\kappa f^{\sigma*}
			\,d\mprime d\thetaprime}
		       {\int_0^1\int_0^1 \varepsilon\,\detJ\, f^\kappa f^{\sigma*}
			\,d\mprime d\thetaprime}~,
	\eeqn
	and similarly for 
	$\langle \fscf \varepsilon\,\,\detJ\, f^\kappa f^{\sigma*}\rangle$,
	where all quantities in the integrands are Dalitz-plot dependent.

	Equation~(\ref{eq:theLikelihood}) invokes the phase 
	space-averaged SCF fraction 
	$\fscfave\equiv\langle\fscf\,\detJ\, f^\kappa f^{\sigma*}\rangle$. 
	The PDF normalization  is decay-dynamics-dependent
	and is computed iteratively. We 
	determine the average SCF fractions separately for each tagging category 
	from MC simulation. 
	
	The width of the dominant $\rho(770)$ resonance is large compared 
	to the mass resolution for TM events (about $8\mevcc$ Gaussian
	resolution). We  therefore neglect resolution effects in the TM 
	model.	
	Mis-reconstructed events	have a poor mass resolution that strongly 
	varies across the Dalitz plot. These events are described in the fit by a 
	two-dimensional resolution function
	\beq
	\label{eq:rscf}
		\Rscf(\mprime_r,\thetaprime_r,\mprime_t,\thetaprime_t)~,
	\eeq
	which represents the probability to reconstruct at the coordinate
	$(\mprime_r,\thetaprime_r)$ an event that has the true coordinate 
	$(\mprime_t,\thetaprime_t)$.  This function obeys the unitary condition
	\beq
		\intl_0^1\intl_0^1 
		\Rscf(\mprime_r,\thetaprime_r,\mprime_t,\thetaprime_t)
		\,d\mprime_r d\thetaprime_r = 1,~
	\eeq
	and is convolved with the signal model. 
	The $\Rscf$ function is obtained from MC simulation.

%	We use the signal model described in Section~\ref{sec:kinmeatics}. 
	The dynamical information in the signal model is described in 
	Section~\ref{sec:kinmeatics} and is connected with $\dt$ via 
	the matrix element in Eq. (\ref{eq:dt}), which serves as the PDF.
	The PDF is modified by the effects of mistagging and the limited vertex 
	resolution~\cite{rhopipaper}. 
	The $\deltat$ resolution function for signal and \B-background 
	events is a sum of three Gaussian distributions, with parameters 
	determined by a fit to fully reconstructed $\Bz$ 
	decays~\cite{BabarS2b}. Since the majority of SCF events 
	arise from mis-reconsructed $\pi^0$ decays which do not 
	affect the vertex resolution, we use the same resolution function for
	TM and SCF events. 
%\\[0.3cm]\noindent

	The Dalitz plot- and $\dt$-dependent PDFs factorize for the 
	charged-\B background modes, but not necessarily
	for the $B^0$ background due to $\BzBzb$ mixing.

 	The charged \B-background
		contribution to the likelihood~(\ref{eq:theLikelihood})
                involves 
		the parameter $\Atag$, multiplied by the tag flavor $\Qtag$ of 
		the event. In the presence of significant ``tag-`charge'' 
		correlation (represented by an effective 
		flavor tag versus Dalitz coordinate correlation),
		it parameterizes possible fake direct \CP violation or 
		asymmetries due to detector effects in these events.
		We also use separate square Dalitz plot PDFs for 
	 	 $B^0$ and $\overline{B}^0$ flavor tags, and a flavor-tag-averaged PDF for 
		untagged events. The PDFs are obtained from MC simulation and are 
		described with the use of non-parametric functions.
		The $\dt$ resolution parameters are determined by a fit to fully 
		reconstructed $\Bp$ decays. For each $\Bp$-background class we obtain 
		effective lifetimes from MC to account for the mis-reconstruction of the 
		event that modifies the nominal $\dt$ resolution function.

	The neutral-$B$ background is parameterized with PDFs that
		depend on the flavor tag of the event. In the case of \CP
		eigenstates, correlations between the flavor tag and the Dalitz 
		coordinate are expected to be small. However, non-\CP  eigenstates,
		such as $a_1^\pm\pi^\mp$, may exhibit such correlations. Both types 
		of decays can have direct
		and mixing-induced \CP  violation. A third type of decay
		involves charged kaons (e.g. $\rho^\pm K^\mp$) 
		and does not exhibit mixing-induced
		\CP  violation, but usually has a strong correlation between the
		flavor tag and the Dalitz plot coordinate, because 
		these decays correspond to  $B$-flavor eigenstates.
		The Dalitz plot PDFs are obtained from MC simulation and are 
		described with the use of non-parametric functions.
		For neutral-$B$ background, the signal $\dt$ resolution model 
		is assumed.

	The Dalitz plot
		treatment of the continuum events is similar to that used
		for charged-$B$ background. 
		The square Dalitz plot PDF for continuum background is 
		obtained from on-resonance events selected in the
		$\mes$ sidebands (defined as $5.225<\mes<5.265$) 
		and corrected for a 5\% feed-through
		from \B decays. A large number of cross checks have been 
		performed to ensure the high fidelity of the empirical shape 
		parameterization. 
		The continuum $\deltat$ distribution is parameterized as the sum of 
		three Gaussian distributions with common mean and 
		three distinct widths.  The widths scale with the estimated $\dt$ uncertainty for each event.
		This yields six shape parameters that are determined by 
		the fit.
 		The model is motivated by the observation that 
		the $\dt$ average is independent of its error, and that the 
		$\dt$ RMS depends linearly on the $\dt$ error.

\subsubsection{PARAMETERIZATION OF THE OTHER VARIABLES}
\label{sec:likmESanddE}

	The $\mes$ distribution of TM signal events is
		parameterized by a bifurcated Crystal Ball function~\cite{PDFsCB},
		which is a combination of a one-sided Gaussian and 
		a Crystal Ball function, given as:
		\beqn
		f(x) = 
		\begin{cases} C e^{(x-m)^2/2s_R^2}   & \hspace{-1.0cm} \text{for $(x-m) > 0$,}
		\\
		        C e^{(x-m)^2/2s_L^2}&\hspace{-1.5cm} \text{for $0 > \frac{x-m}{s_L} > -A$,}
		\\
		       C (\frac{b}{A})^b e^{-\frac{A^2}{2}}\left(\frac{b}{A}-A-\frac{x-m}{s_L}\right)^{-b} &\hspace{-0.3cm} \text{for $\frac{x-m}{s_L} < -A$.}
		\end{cases}
		\eeqn
		The peak position of this function, $m$, 
		is determined by the fit to on-peak data while the other parameters are 
		taken from signal MC. A non-parametric
		function \cite{keys} is used to describe the SCF signal component.

	The $\deprime$ distribution of TM events is
		parameterized by a double Gaussian function, where
		all five parameters depend linearly on $\mpm^2$.
		The parameters of the narrow Gaussian are determined 
		by the fit to data while the others are obtained from 
		signal MC. 
		Mis-reconstructed events are parameterized by a broad
		single Gaussian function whose parameters are taken from signal MC.
		
	Both $\mes$ and $\deprime$ PDFs are parameterized by non-parametric
		functions for all $B$-background classes.  Continuum events are
		parameterized with an Argus shape function~\cite{PDFsArgus}
		\beq
		f(\mes)=C\frac{\mes}{m_{\rm ES}^{\rm max}}\sqrt{1-\left(\frac{\mes}{m_{\rm ES}^{\rm max}}\right)^2} e^{-\xi(1-\left(\frac{\mes}{m_{\rm ES}^{\rm max}}\right)^2)}
		\eeq
		 and 
		a second-order polynomial in $\deprime$,  with parameters 
		determined by the fit.  The value of $m_{ES}^{max}$ is 5.2886 \gevcc.

	We use non-parametric functions to empirically describe the distributions 
		of the NN outputs
		found in the MC simulation for TM and SCF signal events, 
		and for \B-background events. We distinguish tagging categories 
		for TM signal events to account for differences observed in the 
		shapes.
	
	The continuum NN distribution is parameterized by a 
		third-order polynomial.
		The coefficients of the polynomial are determined by the fit.
		Continuum events exhibit a correlation between the Dalitz plot 
		coordinate
		and the inputs to the NN. 
	To account for this correlation, 
		we introduce a linear dependence of the polynomial coefficients
		on the distance of the Dalitz plot coordinate from kinematic 
		boundaries of the Dalitz plot. The parameters describing this
		dependence are determined by the fit.

%% file: bBackground.tex
\setlength{\tabcolsep}{0.0pc}
\begin{tabular*}{\textwidth}{@{\extracolsep{\fill}}llrr}
\hline
Class& Mode								& BR~$[10^{-6}]$		& Expected number of events \\
\hline\\[-0.3cm]
1	& $B^+ \rar \rho^+\rho^0_{\rm\:[long]}$					& $   19.1 \pm\ph{}   3.5$ 	& $   57 \pm\ph{}    11$ \\
1	& $B^+ \rar a_1^+ (\rar (\rho\pi)^+) \piz$ 				& $   20.0 \pm\ph{}   15.0$ 	& $   35 \pm\ph{}   26$ \\
1	& $B^+ \rar a_1^0 (\rar \rho^{\pm} \pi^{\mp}) \pi^+$ 			& $   20.0 \pm        15.0$ 	& $   21 \pm\ph{}    15$ \\
2	& $B^+ \rar \pi^+\rho^0$ 						& $    8.7 \pm\ph{0}   1.0$ 	& $   81 \pm\ph{0}    9$ \\
3	& $B^+ \rar \rho^0 K^+$ 						& $    4.3 \pm\ph{0}   0.6$ 	& $    7 \pm\ph{0}    1$ \\
3	& $B^+ \rar \pi^+K^0_S(\rar\pi^+\pi^-)$ 				& $   8.3 \pm\ph{0}    0.4$ 	& $    11 \pm\ph{0}    1$ \\
4	& $B^+ \rar \pi^0\rho^+$ 						& $   10.8 \pm\ph{0}    1.4$ 	& $   70 \pm\ph{0}    9$ \\
4	& $B^+\rar \pi^+K^0_S(\rar \pi^0\pi^0)$ 				& $    3.7 \pm\ph{0}    0.2$ 	& $   17 \pm\ph{0}    2$ \\
5	& $B^+ \rar \pi^+\pi^0$ 						& $    5.5 \pm\ph{0}    0.6$ 	& $    15 \pm\ph{0}    2$ \\
5	& $B^+ \rar K^+\pi^0$ 							& $   12.1 \pm\ph{0}    0.8$ 	& $    9 \pm\ph{0}    1$ \\
6	& $B^+ \rar (K^{(**)}(1430) \pi)^+ \rar  (\Kp\pi\pi)^+$ 		& $   29.0 \pm\ph{0}    5.4$ 	& $    42 \pm\ph{0}    6$ \\
\hline
7	& $B^0 \rar \pi^- K^{\star +}(\rar K^0_S\pi^+)$ 			& $    3.3 \pm\ph{0}    0.4$ 	& $    2 \pm\ph{0}    1$ \\
8	& $B^0 \rar \rho^+\rho^-_{\rm\:[long]}$ 				& $   25.2 \pm\ph{0}   3.7$ 	& $   74 \pm\ph{}   11$ \\
8	& $B^0 \rar (a_1\pi)^0$ 						& $   39.7 \pm   3.7$ 		& $    43 \pm\ph{}    4$ \\
9	& $B^0 \rar K^+\pi^-$ 							& $   18.9 \pm\ph{0}    0.7$ 	& $    13 \pm\ph{0}   1$ \\
10	& $B^0 \rar \pi^- K^{\star +}(\rar K^+\pi^0)$ 				& $    3.3 \pm\ph{0}    0.4$ 	& $   22 \pm\ph{0}   2$ \\
10	& $B^0 \rar K^{(**)}(1430) \pi \rar K\pi\pi^0$ 				& $   11.2 \pm\ph{0}    2.2$ 	& $    234 \pm\ph{}    37$ \\
11	& $B^0\rar\gamma K^{\star 0}(892,1430)(\rar (K^+\pi^-)^0)$ 		& $   27.4 \pm\ph{0}    1.5$ 	& $    15 \pm\ph{0}    1$ \\
11	& $B^0 \rar \pi^0 K^{\star 0}(\rar K^+\pi^-)$ 				& $    1.3 \pm\ph{0}    0.5$ 	& $   10 \pm\ph{}    4$ \\
11	& $B^0 \rar \eta^\prime(\rar\rho^0\gamma)\pi^0$ 			& $    0.4 \pm\ph{0}    0.2$ 	& $    3 \pm\ph{0}    2$ \\
12	& $B^0 \rar \rho^- K^+$ 						& $    9.9 \pm\ph{0}    1.6$ 	& $   114 \pm\ph{}    17$ \\
13	& $B^0 \rar K^+\pi^-\pi^0_{\rm~[nonres]}$ 				& $    4.6 \pm\ph{0}    4.6$ 	& $   42 \pm\ph{0}    38$ \\
\hline
14	& $B^0 \rar \pi^0K^0_S(\rar\pi^+\pi^-)$ 				& $   5.8 \pm\ph{0}    0.5$ 	& $   55 \pm\ph{0}    4$ \\
15	& $B^0 \rar D^-(\rar\pi^-\pi^0)\pi^+$					& $    7.5 \pm\ph{0}   2.3$ 	& $  661 \pm\ph{}   203$ \\
16	& $B^0 \rar \Dzb(\rar K^+\pi^-)\pi^0$ 					& $   11.0 \pm\ph{0}    3.2$ 	& $   110 \pm\ph{0}    32$ \\
17	& $B^0 \rar \Dzb(\rar \pi^+\pi^-)\pi^0$ 				& $    0.4 \pm\ph{0}    0.1$ 	& $    39 \pm\ph{0}    10$ \\
18	& $B^0 \rar J/\psi(\rar e^+e^-,\mu^+\mu^-)\pi^0$ 			& $    2.6 \pm\ph{0}    0.5$ 	& $   85 \pm\ph{0}    17$ \\
\hline
19	& $B^0 \rar \{\text{neutral generic } b\rar c \text{ decays}\}$ 	& $    -$ 			& $188\pm\ph{0} 30$ \\
20	& $B^+ \rar \{\text{charged generic } b\rar c \text{ decays}\}$ 	& $    -$ 			& $431\pm\ph{} 40$ \\
\hline
\end{tabular*}

%% file: pdfTable.tex
\setlength{\tabcolsep}{0.0pc}
\begin{tabular*}{\textwidth}{@{\extracolsep{\fill}}lcccc}
\hline
 Variable  	& TM Signal	&  SCF Signal  	&   Continuum  	&   $B$-Background  \\
\hline\\[-0.3cm]
$\de$ 		& 	GG	& G		& P2		&	NP	  \\
$\mes$ 		& 	biCB	& NP		& Argus		&	NP	  \\
Neural Net 	& 	NP	& NP		& P3		&	NP	  \\
Dalitz 		& see text	& see text	& NP		&	NP	  \\
$\dt$ 		& 	GGG	& GGG		& GGG		&	GGG	  \\
\hline
\end{tabular*}

%% file: FitResults.tex
\section{FIT RESULTS}
\label{sec:fitResults}

\begin{table*}[t]
\begin{center}
\input{ResultsTable.tex}
\end{center}
\end{table*}

\begin{table*}[t]
\centering
\caption{\label{tab:corrmat}
  Correlation matrix of statistical uncertainties for the $U$
  and $I$ coefficients. 
Since the matrix is symmetric, 
  all elements above the diagonal are omitted.
}
\begin{small}
\input{corrTableStat1Vivace.tex}
\vspace{1.5\baselineskip}
\input{corrTableStat2Vivace.tex}
\end{small}

\end{table*}

\begin{figure*}[p]
   \centerline{  \epsfxsize8.2cm\epsffile{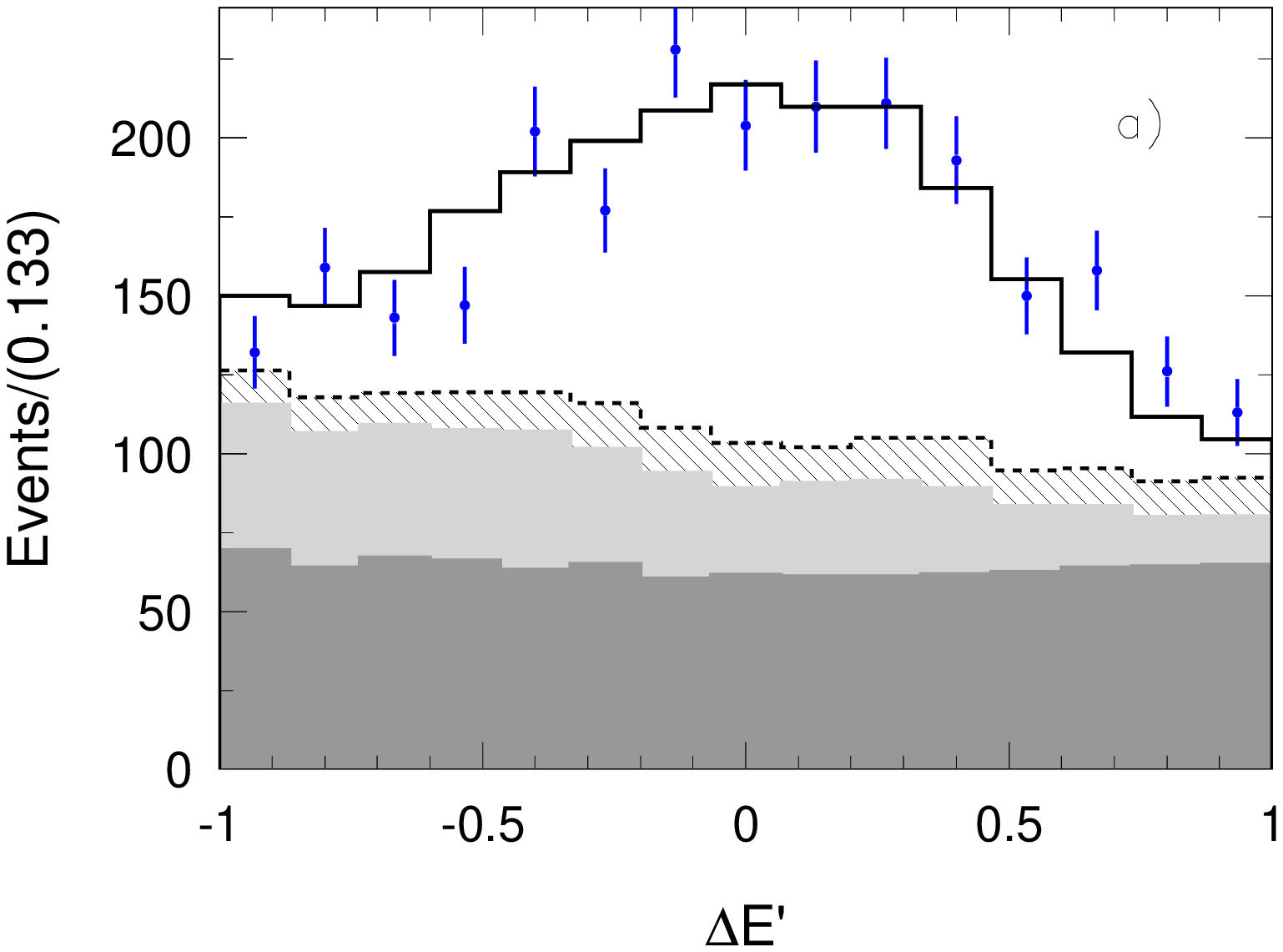}
                \epsfxsize8.2cm\epsffile{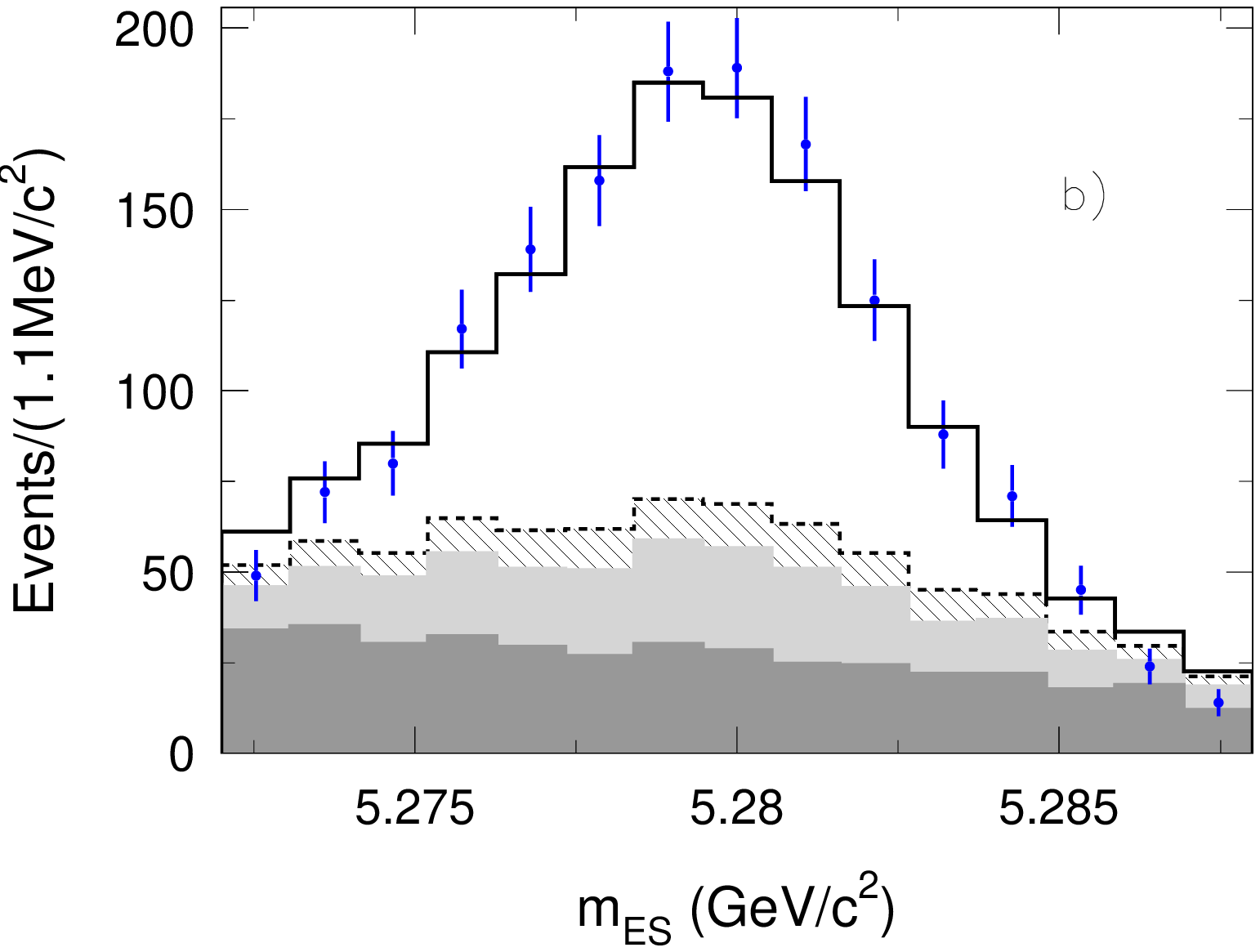}}
  \vspace{-0.1cm}
  \centerline{  \epsfxsize8.2cm\epsffile{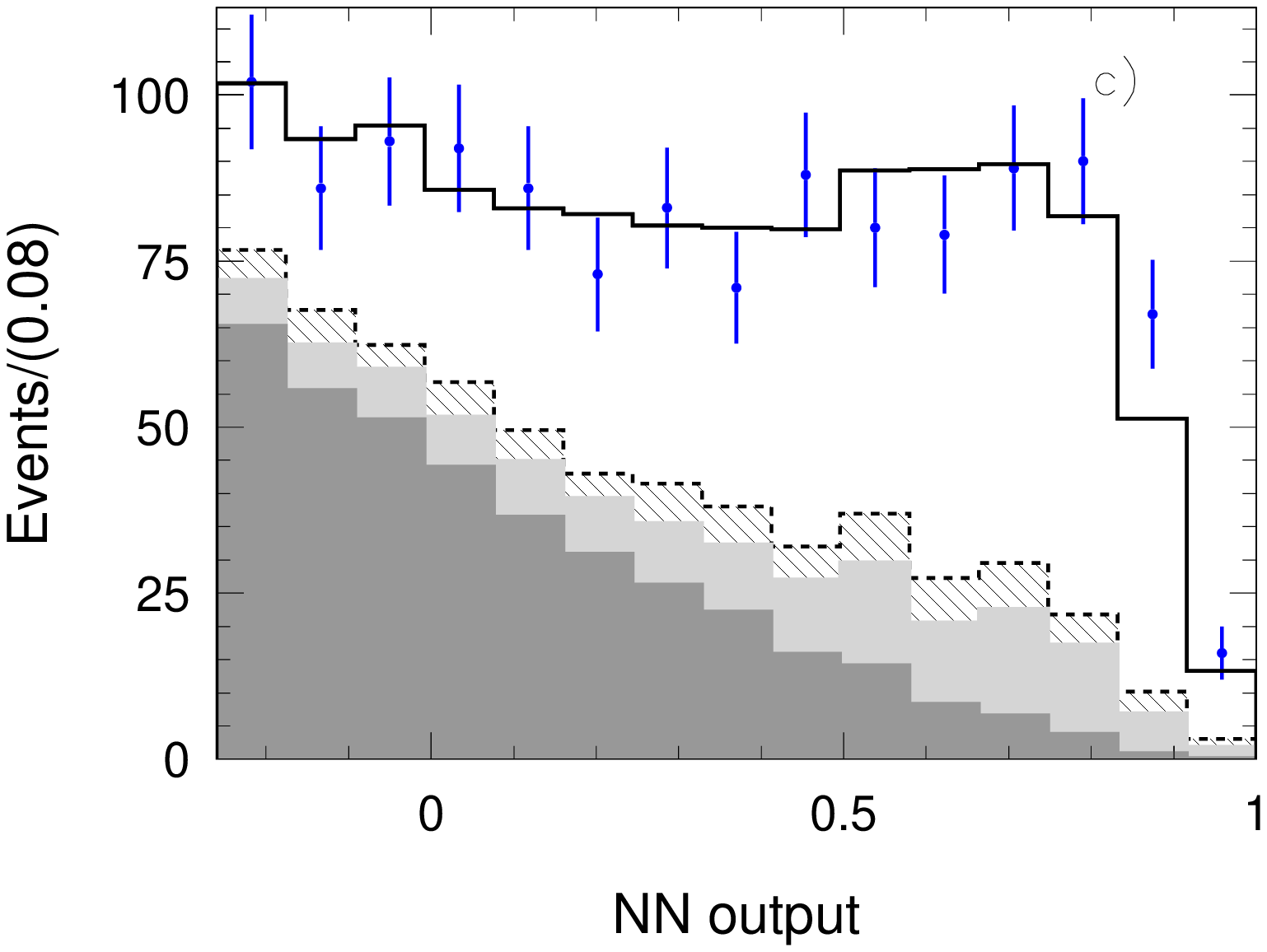} 
                \epsfxsize8.2cm\epsffile{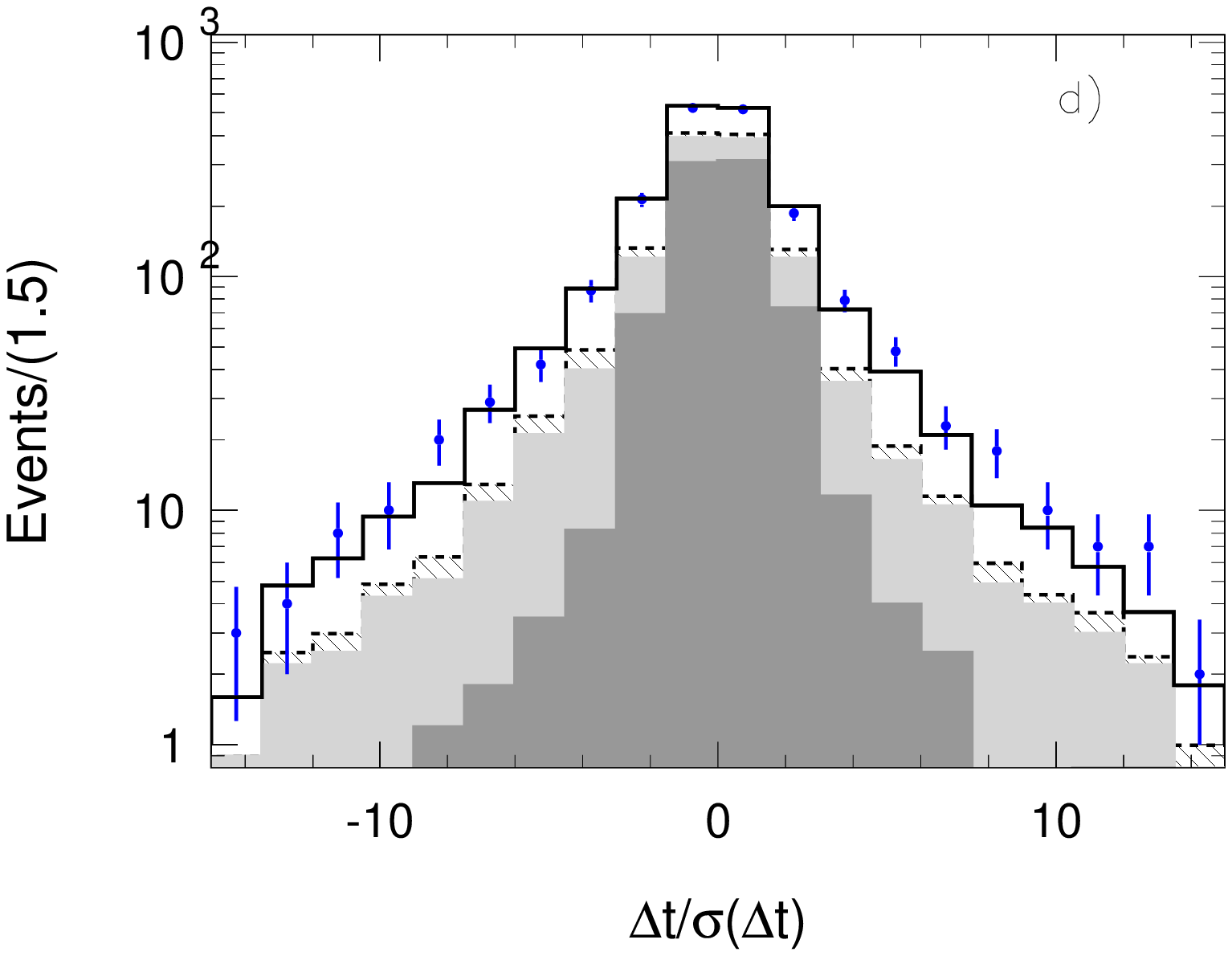}}
  \vspace{-0.1cm}
  \centerline{  \epsfxsize8.2cm\epsffile{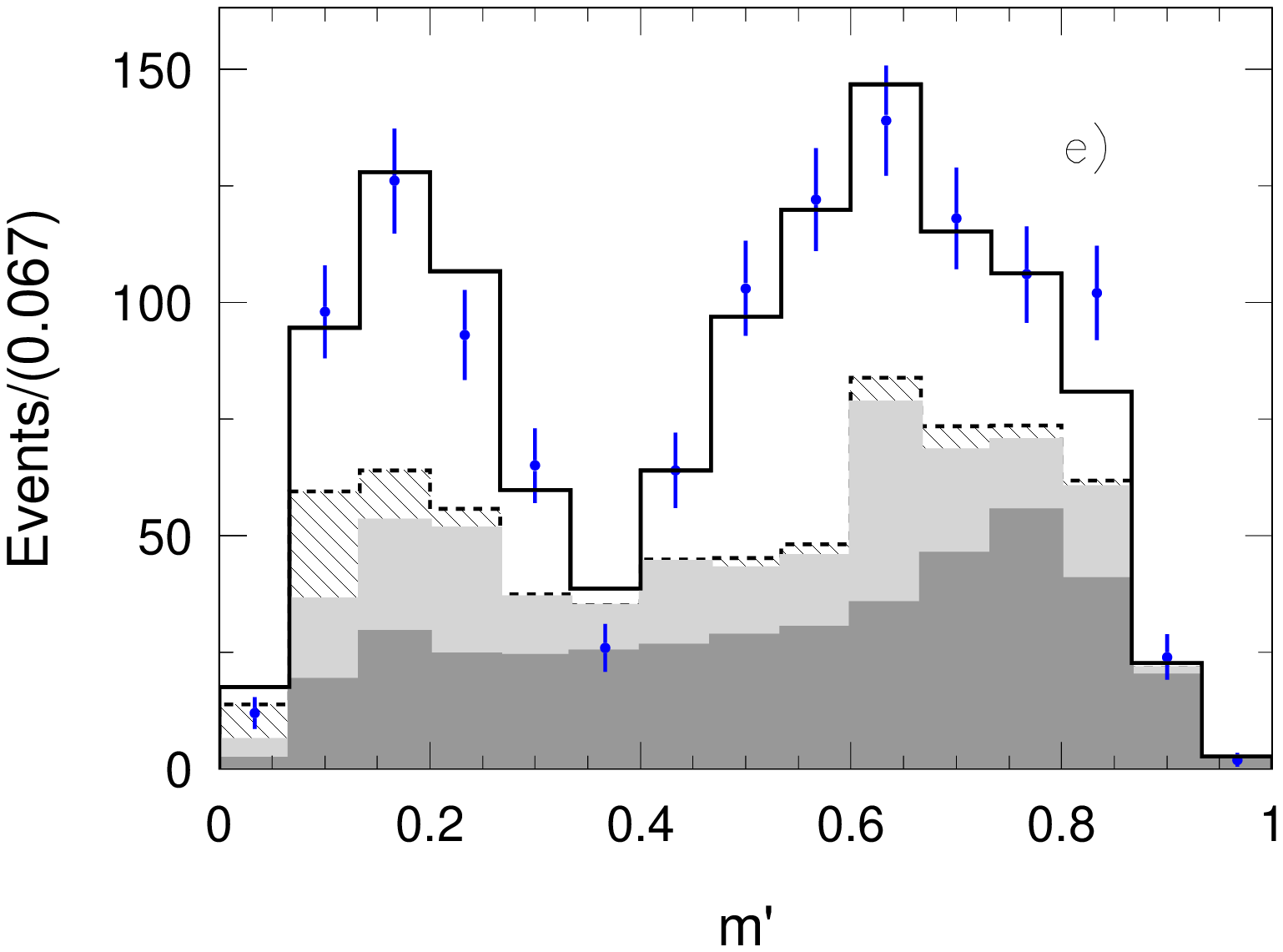}
                \epsfxsize8.2cm\epsffile{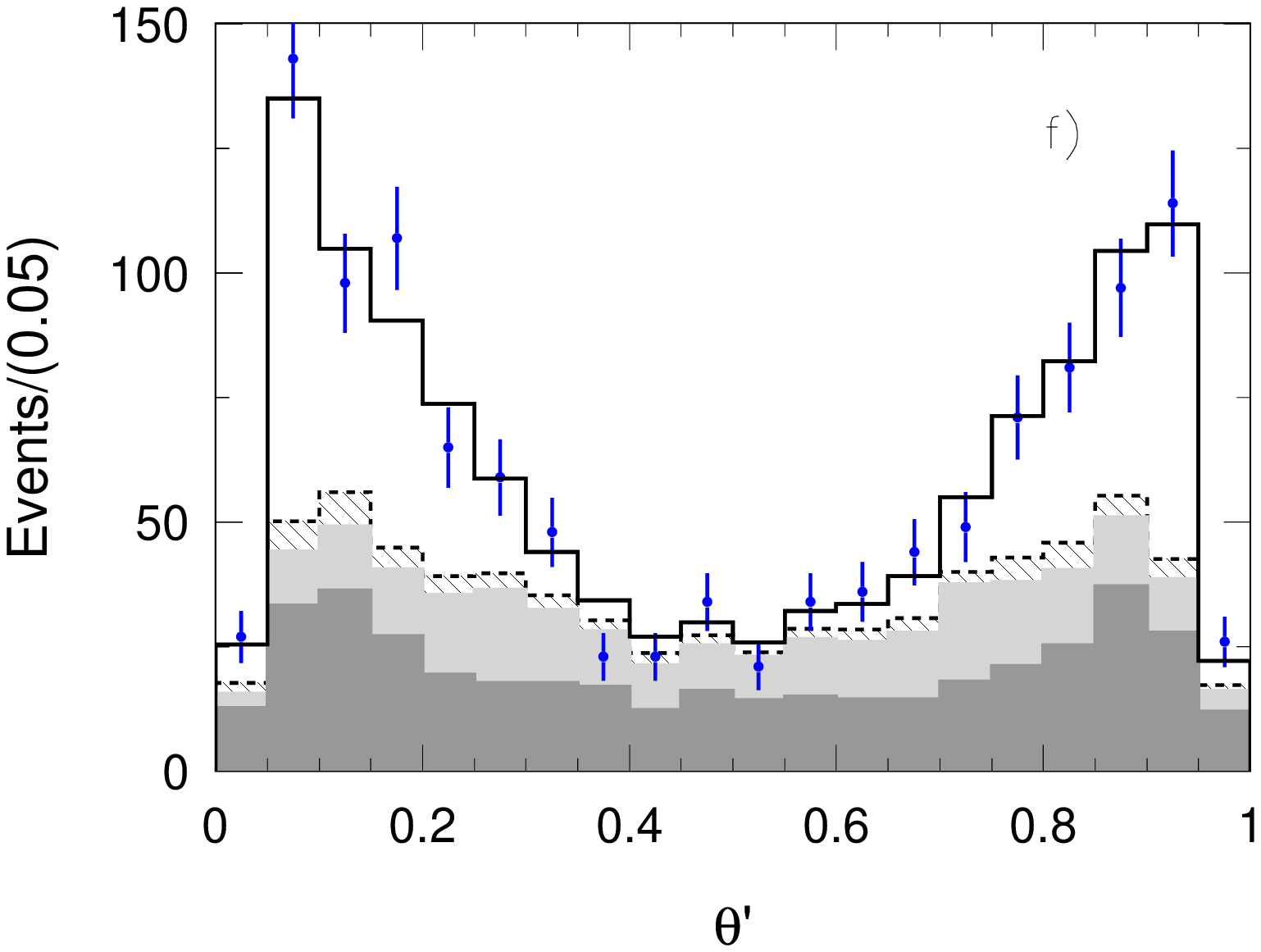}}
 \caption{\label{fig:projections} 
	Distributions of (a-f) $\deprime$, $\mes$, 
	NN output, $\dt/\sigma(\dt)$,  $\mprime$,  and $\thetaprime$ for samples 
	enhanced in $\Btopipipi$ signal. The dots with error-bars correspond to 
	the on-resonance data. The solid histogram shows the
	projection of the fit result. The dark and
	light shaded areas represent the contribution
	from continuum and $B$-background events respectively. 
 	The misreconstructed signal events are represented by the dashed
	histogram. The ratios of signal
	 events over background events in the sample differs for each plot, but are typically $\sim1.2$.}
\end{figure*}

\begin{figure}[t]
  \centerline{\epsfxsize8.2cm\epsffile{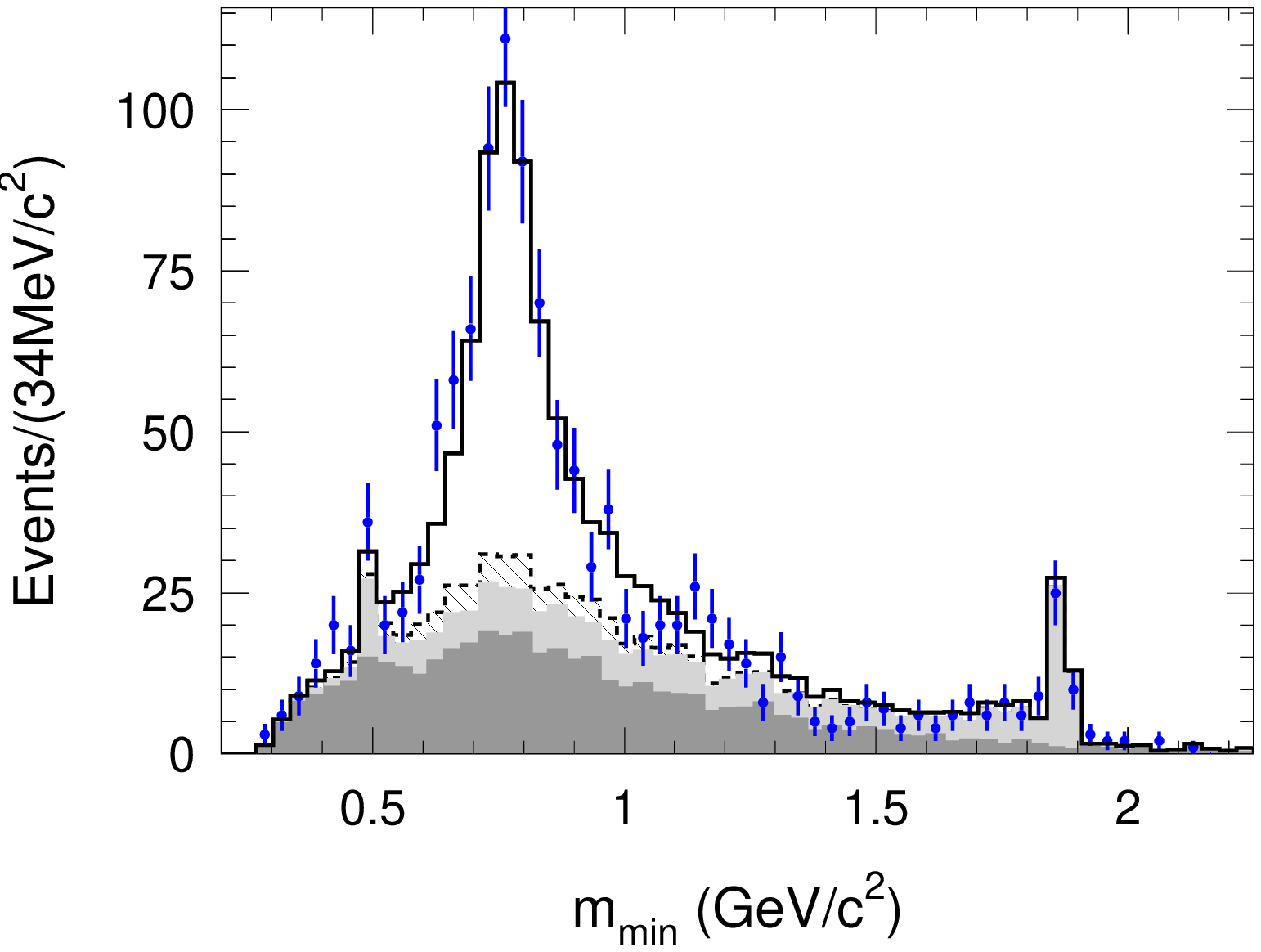}}
  \caption{\label{fig:projection_mmin} 
	Distribution of the minimum of the three di-pion invariant masses,
        for samples enhanced in $\Btopipipi$ signal.  The dots with error-bars correspond to 
	the on-resonance data. The solid histogram shows the
	projection of the fit result. The dark and
	light shaded areas represent the contribution
	from continuum and $B$-background events respectively. 
 	The misreconstructed signal events are represented by the dashed
	histogram.
	 The peaks near 0.5 and 1.8\gevcc are from 
		$B^0\to K_s^0\pi^0$ and $B^0\to D^+\pi^-$ decays, respectively.}
\end{figure}

The maximum-likelihood fit results in a $\Btopipipi$ event yield of
$N_{\tpi}=2067\pm86$, where the error is statistical only. The results 
 for the $U$ and $I$ coefficients are given together with their statistical
and systematic errors in Table~\ref{tab:results}. The corresponding 
correlation matrix is given in Table~\ref{tab:corrmat}. We have 
generated a sample of Monte Carlo experiments to determine the probability
density distributions of the fit parameters. 
Within the statistical 
uncertainties of this sample we find Gaussian distributions 
for the fitted $U$ and $I$ coefficients.  This allows us to 
use the least-squares method to derive other quantities from these
(Section~\ref{sec:Physics}).

The signal is 
dominated by $\Bz\to\rho^\pm\pi^\mp$ decays. We observe an excess of
$\rho^0\piz$ events (see, mainly, $\U_0^+$), which is in agreement with our previous upper
limit~\cite{BABARrho0pi0} and the latest measurement from the Belle 
collaboration~\cite{BELLErho0pi0}.
We find the ratio of $\rho(1450)$/$\rho(770)$ ( $\rho(1700)$/$\rho(770)$ )
rates to be $0.13\pm0.04$ ($0.07\pm0.04$) where the errors are statistically only. 
For the relative strong phase between 
the $\rho(770)$ and  $\rho(1450)$ ($\rho(1700)$) amplitudes we find 
$(163\pm22)^\circ$ $((5\pm36)^\circ$ (statistical errors only), which is 
compatible with the result from $\tau$ and $\epem$ data.
These results for the $\rho$  amplitudes are compatible  with the findings 
in $\tau$ and $\epem$ decays~\cite{taueeref}. 

Figure~\ref{fig:projections} shows distributions 
of $\deprime$, $\mes$, the NN output, and $\dt/\sigma(\dt)$, where $\sigma(\dt)$
is the per-event error on $\dt$, as well as the Dalitz plot 
variables $\mprime$ and $\thetaprime$.  All distributions 
 are enhanced in signal content by selecting on the ratio of 
the probability the event is signal to the total, $P_sig$/$\sum P$, 
excluding the variable plotted. 
Figure~\ref{fig:projection_mmin} shows the distribution
of the minimum of the three di-pion invariant masses, again
enhanced in signal content. This plot shows clearly that
$\rho(770)$ dominates the signal component.

%% file: ResultsTable.tex
\caption{Fit results for the $U$ and $I$ coefficients. 
	The errors given are statistical (first) and systematic (second).
	The free normalization parameter $U_+^+$ is fixed to 1.    The coefficients are defined in Eqn. \ref{eq:UI}.
}
\setlength{\tabcolsep}{0.3pc}
\begin{tabular*}{\textwidth}{@{\extracolsep{\fill}}llc}
\hline
&&\\[-0.3cm]
Parameter   & Description & Result \\[0.15cm]
\hline
&&\\[-0.3cm]
\rule[-1.7mm]{0mm}{5mm}$U_+^+$
     & Coefficient of $|f_+|^2$               & $\phantom{-}1.0$ (fixed)   \\
\rule[-1.7mm]{0mm}{5mm}$U_0^+$
     & Coefficient of $|f_0|^2$               & $\phantom{-}0.28\pm0.07\pm0.04$ \\
\rule[-1.7mm]{0mm}{5mm}$U_-^+$
     & Coefficient of $|f_-|^2$               & $\phantom{-}1.32\pm0.12\pm0.05$  \\[0.15cm]

\rule[-1.7mm]{0mm}{5mm}$U_0^-$
     & Coefficient of $|f_0|^2\cos(\dmd\dmt)$ & $-0.03\pm0.11\pm0.09$  \\
\rule[-1.7mm]{0mm}{5mm}$U_-^-$
     & Coefficient of $|f_-|^2\cos(\dmd\dmt)$ & $-0.32\pm0.14\pm0.05$  \\
\rule[-1.7mm]{0mm}{5mm}$U_+^-$
     & Coefficient of $|f_+|^2\cos(\dmd\dmt)$ & $\phantom{-}0.54\pm0.15\pm0.05$  \\[0.15cm]

\rule[-1.7mm]{0mm}{5mm}$I_0$
     & Coefficient of $|f_0|^2\sin(\dmd\dmt)$ & $\phantom{-} 0.01\pm0.06\pm0.01$ \\
\rule[-1.7mm]{0mm}{5mm}$I_-$
     & Coefficient of $|f_-|^2\sin(\dmd\dmt)$ & $- 0.01\pm0.10\pm0.02$ \\
\rule[-1.7mm]{0mm}{5mm}$I_+$
     & Coefficient of $|f_+|^2\sin(\dmd\dmt)$ & $- 0.02\pm0.10\pm0.03$\\[0.15cm]

\rule[-1.7mm]{0mm}{5mm}$U_{+-}^{+,\I}$
     & Coefficient of $\I[f_+f_-^*]$ & 		$-0.07\pm0.71\pm0.73$   \\
\rule[-1.7mm]{0mm}{5mm}$U_{+-}^{+,\R}$
     & Coefficient of $\R[f_+f_-^*]$ & 		$\phantom{-}0.17\pm0.49\pm0.31$    \\
\rule[-1.7mm]{0mm}{5mm}$U_{+-}^{-,\I}$
     & Coefficient of $\I[f_+f_-^*]\cos(\dmd\dmt)$ & $-0.38\pm1.06\pm0.36$  \\
\rule[-1.7mm]{0mm}{5mm}$U_{+-}^{-,\R}$
     & Coefficient of $\R[f_+f_-^*]\cos(\dmd\dmt)$ & $\phantom{-}2.23\pm1.00\pm0.43$   \\
\rule[-1.7mm]{0mm}{5mm}$I_{+-}^{\I}$
     & Coefficient of $\I[f_+f_-^*]\sin(\dmd\dmt)$ & $-1.99\pm1.25\pm0.34$  \\
\rule[-1.7mm]{0mm}{5mm}$I_{+-}^{\R}$
     & Coefficient of $\R[f_+f_-^*]\sin(\dmd\dmt)$ & $\phantom{-}1.90\pm2.03\pm0.65$\\[0.15cm]

\rule[-1.7mm]{0mm}{5mm}$U_{+0}^{+,\I}$
     & Coefficient of $\I[f_+f_0^*]$ & 		$-0.16\pm0.57\pm0.14$   \\
\rule[-1.7mm]{0mm}{5mm}$U_{+0}^{+,\R}$
     & Coefficient of $\R[f_+f_0^*]$ & 		$-1.08\pm0.48\pm0.20$    \\
\rule[-1.7mm]{0mm}{5mm}$U_{+0}^{-,\I}$
     & Coefficient of $\I[f_+f_0^*]\cos(\dmd\dmt)$ & $-1.66\pm0.94\pm0.25$  \\
\rule[-1.7mm]{0mm}{5mm}$U_{+0}^{-,\R}$
     & Coefficient of $\R[f_+f_0^*]\cos(\dmd\dmt)$ & $-0.18\pm0.88\pm0.35$ \\
\rule[-1.7mm]{0mm}{5mm}$I_{+0}^{\I}$
     & Coefficient of $\I[f_+f_0^*]\sin(\dmd\dmt)$ & $-0.21\pm1.06\pm0.25$  \\
\rule[-1.7mm]{0mm}{5mm}$I_{+0}^{\R}$
     & Coefficient of $\R[f_+f_0^*]\sin(\dmd\dmt)$ & $\phantom{-}0.41\pm1.30\pm0.41$ \\[0.15cm]

\rule[-1.7mm]{0mm}{5mm}$U_{-0}^{+,\I}$
     & Coefficient of $\I[f_-f_0^*]$ & 		$-0.17\pm0.50\pm0.23$ \\
\rule[-1.7mm]{0mm}{5mm}$U_{-0}^{+,\R}$
     & Coefficient of $\R[f_-f_0^*]$ & 		$-0.36\pm0.38\pm0.08$    \\
\rule[-1.7mm]{0mm}{5mm}$U_{-0}^{-,\I}$
     & Coefficient of $\I[f_-f_0^*]\cos(\dmd\dmt)$ & $\phantom{-}0.12\pm0.75\pm0.22$  \\
\rule[-1.7mm]{0mm}{5mm}$U_{-0}^{-,\R}$
     & Coefficient of $\R[f_-f_0^*]\cos(\dmd\dmt)$ & $-0.63\pm0.72\pm0.32$   \\
\rule[-1.7mm]{0mm}{5mm}$I_{-0}^{\I}$
     & Coefficient of $\I[f_-f_0^*]\sin(\dmd\dmt)$ & $\phantom{-}1.23\pm1.07\pm0.29$  \\
\rule[-1.7mm]{0mm}{5mm}$I_{-0}^{\R}$
     & Coefficient of $\R[f_-f_0^*]\sin(\dmd\dmt)$ & $\phantom{-}0.41\pm1.30\pm0.21$ \\[0.15cm]
\hline
\end{tabular*}
\label{tab:results}

%% file: corrTableStat1Vivace.tex
\setlength{\tabcolsep}{0.0pc}
\begin{tabular*}{\textwidth}{@{\extracolsep{\fill}}lccccccccccccc}
\hline
 && \\[-0.3cm]
\rule[-6pt]{0pt}{18pt}  & $\sigPiNb$  &      $\Iz$  &      $\IM$  &   $\IMzIm$  &   $\IMzRe$  &      $\IP$  &   $\IPzIm$  &   $\IPzRe$  &   $\IPMIm$  &   $\IPMRe$  &     $\Uzm$  &     $\Uzp$  &  $\UMzmIm$ \\[0.15cm]
\hline
 &&\\[-0.3cm] $\sigPiNb$  &  $\phantom{-}1.00$  &  &  &  &  &  &  &  &  &  &  &  & \\
      $\Iz$  &  $-0.02$  &  $\phantom{-}1.00$  &  &  &  &  &  &  &  &  &  &  & \\
      $\IM$  &  $-0.04$  &  $-0.04$  &  $\phantom{-}1.00$  &  &  &  &  &  &  &  &  &  & \\
   $\IMzIm$  &  $-0.09$  &  $-0.11$  &  $\phantom{-}0.28$  &  $\phantom{-}1.00$  &  &  &  &  &  &  &  &  & \\
   $\IMzRe$  &  $-0.03$  &  $\phantom{-}0.28$  &  $-0.18$  &  $-0.15$  &  $\phantom{-}1.00$  &  &  &  &  &  &  &  & \\
      $\IP$  &  $\phantom{-}0.06$  &  $-0.04$  &  $-0.20$  &  $-0.21$  &  $\phantom{-}0.17$  &  $\phantom{-}1.00$  &  &  &  &  &  &  & \\
   $\IPzIm$  &  $\phantom{-}0.06$  &  $-0.03$  &  $-0.11$  &  $-0.14$  &  $\phantom{-}0.09$  &  $\phantom{-}0.38$  &  $\phantom{-}1.00$  &  &  &  &  &  & \\
   $\IPzRe$  &  $-0.17$  &  $\phantom{-}0.30$  &  $\phantom{-}0.18$  &  $\phantom{-}0.17$  &  $-0.06$  &  $-0.35$  &  $-0.45$  &  $\phantom{-}1.00$  &  &  &  &  & \\
   $\IPMIm$  &  $\phantom{-}0.09$  &  $\phantom{-}0.11$  &  $\phantom{-}0.14$  &  $-0.17$  &  $\phantom{-}0.10$  &  $\phantom{-}0.21$  &  $\phantom{-}0.11$  &  $-0.03$  &  $\phantom{-}1.00$  &  &  &  & \\
   $\IPMRe$  &  $-0.24$  &  $\phantom{-}0.04$  &  $\phantom{-}0.36$  &  $\phantom{-}0.28$  &  $-0.15$  &  $-0.46$  &  $-0.25$  &  $\phantom{-}0.43$  &  $-0.01$  &  $\phantom{-}1.00$  &  &  & \\
     $\Uzm$  &  $-0.03$  &  $\phantom{-}0.07$  &  $\phantom{-}0.08$  &  $-0.05$  &  $-0.11$  &  $-0.06$  &  $-0.02$  &  $\phantom{-}0.13$  &  $\phantom{-}0.09$  &  $\phantom{-}0.15$  &  $\phantom{-}1.00$  &  & \\
     $\Uzp$  &  $\phantom{-}0.04$  &  $-0.02$  &  $\phantom{-}0.20$  &  $\phantom{-}0.32$  &  $-0.19$  &  $-0.24$  &  $-0.19$  &  $\phantom{-}0.30$  &  $-0.13$  &  $\phantom{-}0.35$  &  $\phantom{-}0.11$  &  $\phantom{-}1.00$  & \\
  $\UMzmIm$  &  $\phantom{-}0.01$  &  $\phantom{-}0.13$  &  $-0.11$  &  $-0.14$  &  $\phantom{-}0.42$  &  $\phantom{-}0.14$  &  $\phantom{-}0.08$  &  $-0.07$  &  $\phantom{-}0.10$  &  $-0.13$  &  $-0.36$  &  $-0.18$  &  $\phantom{-}1.00$ \\
  $\UMzmRe$  &  $\phantom{-}0.02$  &  $\phantom{-}0.07$  &  $-0.05$  &  $-0.44$  &  $\phantom{-}0.03$  &  $\phantom{-}0.06$  &  $\phantom{-}0.05$  &  $\phantom{-}0.01$  &  $\phantom{-}0.17$  &  $-0.06$  &  $\phantom{-}0.04$  &  $-0.19$  &  $\phantom{-}0.13$ \\
  $\UMzpIm$  &  $\phantom{-}0.07$  &  $\phantom{-}0.18$  &  $-0.14$  &  $-0.39$  &  $\phantom{-}0.21$  &  $\phantom{-}0.22$  &  $\phantom{-}0.18$  &  $-0.14$  &  $\phantom{-}0.27$  &  $-0.26$  &  $\phantom{-}0.03$  &  $-0.56$  &  $\phantom{-}0.31$ \\
  $\UMzpRe$  &  $-0.05$  &  $\phantom{-}0.16$  &  $\phantom{-}0.07$  &  $-0.21$  &  $\phantom{-}0.07$  &  $-0.01$  &  $-0.01$  &  $\phantom{-}0.14$  &  $\phantom{-}0.25$  &  $\phantom{-}0.08$  &  $\phantom{-}0.09$  &  $-0.05$  &  $\phantom{-}0.19$ \\
     $\UMm$  &  $\phantom{-}0.11$  &  $-0.01$  &  $-0.03$  &  $-0.06$  &  $\phantom{-}0.01$  &  $\phantom{-}0.01$  &  $\phantom{-}0.02$  &  $-0.06$  &  $\phantom{-}0.01$  &  $-0.08$  &  $-0.09$  &  $-0.08$  &  $\phantom{-}0.12$ \\
     $\UMp$  &  $-0.12$  &  $\phantom{-}0.14$  &  $\phantom{-}0.08$  &  $-0.02$  &  $\phantom{-}0.08$  &  $-0.06$  &  $-0.06$  &  $\phantom{-}0.22$  &  $\phantom{-}0.06$  &  $\phantom{-}0.20$  &  $\phantom{-}0.11$  &  $\phantom{-}0.18$  &  $\phantom{-}0.10$ \\
  $\UPzmIm$  &  $\phantom{-}0.26$  &  $\phantom{-}0.03$  &  $-0.17$  &  $-0.19$  &  $\phantom{-}0.10$  &  $\phantom{-}0.17$  &  $\phantom{-}0.11$  &  $-0.20$  &  $\phantom{-}0.08$  &  $-0.40$  &  $-0.26$  &  $-0.31$  &  $\phantom{-}0.15$ \\
  $\UPzmRe$  &  $\phantom{-}0.13$  &  $-0.02$  &  $\phantom{-}0.00$  &  $\phantom{-}0.08$  &  $-0.18$  &  $-0.13$  &  $-0.23$  &  $\phantom{-}0.07$  &  $-0.05$  &  $-0.05$  &  $-0.04$  &  $\phantom{-}0.03$  &  $-0.11$ \\
  $\UPzpIm$  &  $\phantom{-}0.03$  &  $\phantom{-}0.12$  &  $-0.17$  &  $-0.41$  &  $\phantom{-}0.36$  &  $\phantom{-}0.34$  &  $\phantom{-}0.34$  &  $-0.31$  &  $\phantom{-}0.25$  &  $-0.29$  &  $-0.02$  &  $-0.54$  &  $\phantom{-}0.29$ \\
  $\UPzpRe$  &  $\phantom{-}0.23$  &  $-0.03$  &  $-0.16$  &  $-0.25$  &  $\phantom{-}0.13$  &  $\phantom{-}0.25$  &  $\phantom{-}0.28$  &  $-0.49$  &  $\phantom{-}0.16$  &  $-0.44$  &  $-0.12$  &  $-0.36$  &  $\phantom{-}0.12$ \\
  $\UPMmIm$  &  $-0.14$  &  $-0.04$  &  $\phantom{-}0.03$  &  $\phantom{-}0.15$  &  $-0.04$  &  $-0.08$  &  $-0.07$  &  $\phantom{-}0.09$  &  $-0.11$  &  $\phantom{-}0.19$  &  $\phantom{-}0.01$  &  $\phantom{-}0.13$  &  $-0.08$ \\
  $\UPMmRe$  &  $-0.12$  &  $-0.10$  &  $-0.05$  &  $\phantom{-}0.14$  &  $-0.05$  &  $-0.01$  &  $-0.04$  &  $-0.01$  &  $-0.20$  &  $\phantom{-}0.07$  &  $-0.10$  &  $\phantom{-}0.09$  &  $-0.04$ \\
  $\UPMpIm$  &  $\phantom{-}0.12$  &  $-0.20$  &  $-0.09$  &  $\phantom{-}0.18$  &  $-0.21$  &  $-0.03$  &  $-0.01$  &  $-0.17$  &  $-0.18$  &  $-0.22$  &  $-0.17$  &  $\phantom{-}0.02$  &  $-0.17$ \\
  $\UPMpRe$  &  $-0.15$  &  $\phantom{-}0.05$  &  $\phantom{-}0.07$  &  $\phantom{-}0.01$  &  $\phantom{-}0.06$  &  $\phantom{-}0.00$  &  $-0.03$  &  $\phantom{-}0.12$  &  $\phantom{-}0.22$  &  $\phantom{-}0.13$  &  $\phantom{-}0.10$  &  $\phantom{-}0.10$  &  $\phantom{-}0.03$ \\
     $\UPm$  &  $-0.05$  &  $-0.01$  &  $\phantom{-}0.00$  &  $\phantom{-}0.05$  &  $-0.06$  &  $-0.07$  &  $-0.05$  &  $\phantom{-}0.05$  &  $-0.03$  &  $\phantom{-}0.04$  &  $-0.03$  &  $\phantom{-}0.05$  &  $-0.04$ \\
[0.15cm]\hline
\end{tabular*}

%% file: corrTableStat2Vivace.tex
\setlength{\tabcolsep}{0.0pc}
\begin{tabular*}{\textwidth}{@{\extracolsep{\fill}}lcccccccccccccc}
\hline
 && \\[-0.3cm]
\rule[-6pt]{0pt}{18pt}  &  $\UMzmRe$  &  $\UMzpIm$  &  $\UMzpRe$  &     $\UMm$  &     $\UMp$  &  $\UPzmIm$  &  $\UPzmRe$  &  $\UPzpIm$  &  $\UPzpRe$  &  $\UPMmIm$  &  $\UPMmRe$  &  $\UPMpIm$  &  $\UPMpRe$  &     $\UPm$ \\[0.15cm]
\hline
 &&\\[-0.3cm]  $\UMzmRe$  &  $\phantom{-}1.00$  &  &  &  &  &  &  &  &  &  &  &  &  & \\
  $\UMzpIm$  &  $\phantom{-}0.36$  &  $\phantom{-}1.00$  &  &  &  &  &  &  &  &  &  &  &  & \\
  $\UMzpRe$  &  $\phantom{-}0.34$  &  $\phantom{-}0.25$  &  $\phantom{-}1.00$  &  &  &  &  &  &  &  &  &  &  & \\
     $\UMm$  &  $\phantom{-}0.19$  &  $\phantom{-}0.06$  &  $\phantom{-}0.03$  &  $\phantom{-}1.00$  &  &  &  &  &  &  &  &  &  & \\
     $\UMp$  &  $\phantom{-}0.11$  &  $\phantom{-}0.20$  &  $\phantom{-}0.28$  &  $-0.13$  &  $\phantom{-}1.00$  &  &  &  &  &  &  &  &  & \\
  $\UPzmIm$  &  $\phantom{-}0.07$  &  $\phantom{-}0.22$  &  $-0.04$  &  $\phantom{-}0.10$  &  $-0.16$  &  $\phantom{-}1.00$  &  &  &  &  &  &  &  & \\
  $\UPzmRe$  &  $-0.01$  &  $-0.08$  &  $-0.05$  &  $\phantom{-}0.07$  &  $-0.05$  &  $\phantom{-}0.32$  &  $\phantom{-}1.00$  &  &  &  &  &  &  & \\
  $\UPzpIm$  &  $\phantom{-}0.22$  &  $\phantom{-}0.56$  &  $\phantom{-}0.17$  &  $-0.00$  &  $\phantom{-}0.05$  &  $\phantom{-}0.20$  &  $-0.25$  &  $\phantom{-}1.00$  &  &  &  &  &  & \\
  $\UPzpRe$  &  $\phantom{-}0.08$  &  $\phantom{-}0.23$  &  $\phantom{-}0.02$  &  $\phantom{-}0.11$  &  $-0.31$  &  $\phantom{-}0.37$  &  $-0.06$  &  $\phantom{-}0.41$  &  $\phantom{-}1.00$  &  &  &  &  & \\
  $\UPMmIm$  &  $-0.15$  &  $-0.20$  &  $-0.11$  &  $-0.27$  &  $-0.05$  &  $-0.15$  &  $-0.03$  &  $-0.12$  &  $-0.17$  &  $\phantom{-}1.00$  &  &  &  & \\
  $\UPMmRe$  &  $-0.12$  &  $-0.19$  &  $-0.21$  &  $\phantom{-}0.08$  &  $\phantom{-}0.02$  &  $-0.07$  &  $\phantom{-}0.03$  &  $-0.16$  &  $-0.16$  &  $\phantom{-}0.09$  &  $\phantom{-}1.00$  &  &  & \\
  $\UPMpIm$  &  $-0.22$  &  $-0.33$  &  $-0.35$  &  $\phantom{-}0.06$  &  $-0.48$  &  $\phantom{-}0.15$  &  $\phantom{-}0.22$  &  $-0.33$  &  $\phantom{-}0.12$  &  $\phantom{-}0.10$  &  $\phantom{-}0.12$  &  $\phantom{-}1.00$  &  & \\
  $\UPMpRe$  &  $\phantom{-}0.03$  &  $-0.04$  &  $\phantom{-}0.17$  &  $-0.02$  &  $\phantom{-}0.10$  &  $-0.19$  &  $-0.19$  &  $\phantom{-}0.05$  &  $-0.07$  &  $\phantom{-}0.03$  &  $-0.15$  &  $-0.24$  &  $\phantom{-}1.00$  & \\
     $\UPm$  &  $-0.02$  &  $-0.03$  &  $-0.03$  &  $-0.07$  &  $\phantom{-}0.11$  &  $\phantom{-}0.14$  &  $\phantom{-}0.30$  &  $-0.11$  &  $-0.11$  &  $\phantom{-}0.18$  &  $\phantom{-}0.20$  &  $\phantom{-}0.02$  &  $-0.12$  &  $\phantom{-}1.00$ \\
[0.15cm]\hline
\end{tabular*}

%% file: Systematics.tex
\section{SYSTEMATIC STUDIES}
\label{sec:Systematics}

\begin{table*}[t]
\begin{center}
\caption{ \label{tab:systematics}
        Summary of systematic uncertainties.}
\input{sysTable.tex}
\vspace{-0.2cm}
\end{center}
\end{table*}

\begin{table*}[t]
\centering
\caption{\label{tab:corrmatSyst}
  Correlation matrix of systematic uncertainties for the $U$
  and $I$ coefficients. 
	Since the matrix is symmetric, 
  all elements above the diagonal are omitted.
}
\begin{small}
\input{corrTableSyst1Vivace.tex}
\vspace{1.5\baselineskip}
\input{corrTableSyst2Vivace.tex}
\end{small}
\label{tab:sysCorrelationsFinal}
\end{table*}

The contributions to the systematic error on the signal parameters are 
summarized in Table~\ref{tab:systematics}. Table~\ref{tab:corrmatSyst}
summarizes the correlation coefficients extracted from the
systematic covariance matrix.  For a given systematic effect, we vary 
a parameter in the fit (e.g. the $\rho(770)$ mass), refit the data, and 
construct the systematic covariance matrix for that source based on the deviations
of the $U$ and $I$ coefficients from the nominal values.  The $(i,j)$  matrix element is given as
\beqn
	s_{i,j} &\equiv&	\delta_i\delta_j\\[0.3cm]
\eeqn	
where $\delta_i$ is the difference between the two two fits for variable $i$.  
The total systematic covariance matrix is
 obtained by adding together the covariance matrices in quadrature from the
different systematic sources.  

To estimate the contribution to $\Btopipipi$ decay from other
resonances and non-resonant decays, we fit the on-peak data including
 these other possible decays in the fit model.
For simplicity, we assume a uniform Dalitz distribution
for the non-resonant events and consider possible non-$\rho$ resonances
including $f_0(980)$, $f_2(1270)$, and a low mass $S$-wave $\sigma$ whose
mass and width we take to be 478\mevcc and 324\mevcc, respectively \cite{PDG}.
The fit does not find a significant signal for any of those decays.
However, the inclusion of the broad, low mass $\pipi$ $S$-wave
component significantly degrades our ability to identify $\rho^0\piz$ events.
The systematic effect (contained in the ``Dalitz plot model'' rows in 
Table~\ref{tab:systematics})
is estimated by generating Monte Carlo samples including the other $\Btopipipi$ modes
and fitting  with the nominal setup, where only $\rho(770)$ is taken
into account.

We vary the mass and width of the $\rho(770)$, $\rho(1450)$, and $\rho(1700)$ resonances
within ranges that 
exceed twice the errors found for these parameters extracted from 
$\tau$ decays and $\epem$ annihilations~\cite{taueeref}, and assign the observed
shifts in the measured $U$ and $I$ coefficients as systematic uncertainties
(``$\rho,\rho',\rho''$ lineshape'' in Table~\ref{tab:systematics}). 
Since some of the $U$ and 
$I$ coefficients exhibit significant dependence on the $\rho(1450)$ 
and $\rho(1700)$ contributions, we leave their amplitudes (phases and 
fractions) free to vary in all fits.

To validate the fitting tool, we perform fits on large MC samples with
the measured proportions of signal, continuum, and $B$-background events.
No significant biases are observed in these fits. The statistical
uncertainties on their fit parameters are taken as systematic uncertainties
(``Fit bias'' in Table~\ref{tab:systematics}).

Another potentially large source of systematic uncertainty is the $B$-background model. 
The expected event yields from the background modes are varied according 
to the uncertainties in the measured or estimated branching fractions
(``$N_{\rm{Background}}$'' in Table~\ref{tab:systematics}).  
Since $B$-background modes may exhibit  \CP violation, the corresponding 
parameters are varied either within their measured ranges (if available) 
or within $\pm 0.5$ (if unmeasured)
(``$B$ background \CP'' in Table~\ref{tab:systematics}). 

Other systematic effects are much less important to the measurements
of $U$ and $I$ coefficients and  are combined in 
the ``Others'' field in Table~\ref{tab:systematics}. Details are given
below.

The parameters for 
the continuum events are determined by the fit. No additional systematic 
uncertainties are assigned to them. An exception to this is the Dalitz 
plot PDF; to estimate the systematic
uncertainty from the \mes sideband extrapolation, we select large 
samples of off-resonance data by loosening the requirements on \de and 
the NN. We compare the distributions of $\mprime$ and $\thetaprime$ 
between the \mes sideband and the signal region. No significant 
differences are found. We assign as systematic error the effect seen when
weighting the continuum Dalitz plot PDF by the ratio of the 2-dimensioal
histograms taken from the signal region
and sideband data sets. This effect is mostly statistical in origin. 

The uncertainties associated with $\dmd$ and $\tau$ are
estimated by varying these parameters within the uncertainties
on the world averages~\cite{PDG}.

The systematic effects due to the signal PDFs 
comprise uncertainties in the
PDF parameterization, the treatment of misreconstructed decays, the
tagging performance, and the modeling of the signal contributions.

When the signal PDFs are determined from fits to a control sample
of fully reconstructed \B decays to exclusive final states with
charm, the uncertainties are obtained by varying the parameters
within the statistical uncertainties.
In other cases, the dominant parameters have been
left free to vary in the fit, and the differences observed in these
fits are taken as systematic errors.

The average fraction of misreconstructed signal events predicted by the MC
simulation has been verified with fully reconstructed $\B\to D\rho$
events~\cite{rhopipaper}. No significant differences between data and
the simulation are found. We vary $\fscfave$ for all tagging categories
relatively by $25\%$ to estimate the systematic uncertainty.

  As is done for the signal PDFs, we vary the $\dt$ resolution parameters and
the flavor-tagging parameters within their uncertainties and assign
the differences observed in the  data fit with respect to the 
nominal fit as systematic errors.

The systematic errors for the parameters that measure interference
effects are dominated by the uncertainty in the signal model, mainly
the  description of the $\rho$ resonance tails. For the other parameters,
the uncertainty on the fit bias and the \B-background contamination 
are important.

As a validation of our treatment of the time dependence  we allow
$\tau_{\Bz}$ to vary in the fit. We find
$\tau_{\Bz} = (1.513\pm 0.066)\ps$,
while the remaining free parameters are consistent with the nominal fit.
To validate the SCF modeling, we leave the average SCF fractions per tagging
category free to vary in the fit and find results that are consistent
with the MC prediction. 

%% file: sysTable.tex
\setlength{\tabcolsep}{0.0pc}
\begin{tabular*}{\textwidth}{@{\extracolsep{\fill}}lcccccc}
\hline
 \rule[-6pt]{0pt}{18pt}  &      $\Iz$  &      $\IM$  &   $\IMzIm$  &   $\IMzRe$  &      $\IP$  &   $\IPzIm$  \\
\hline\\[-0.35cm]
 Dalitz plot model  &  0.010  &  0.006  &  0.110  &  0.102  &  0.020  &  0.018  \\
 $\rho$,$\rho'$,$\rho''$ lineshape   &  0.003  &  0.012  &  0.240  &  0.103  &  0.009  &  0.225  \\
  Fit bias  &  0.000  &  0.002  &  0.012  &  0.049  &  0.005  &  0.015  \\
$N_{\rm Background}$  &  0.005  &  0.005  &  0.072  &  0.096  &  0.005  &  0.045  \\
    \B background \CP  &  0.004  &  0.013  &  0.064  &  0.083  &  0.009  &  0.050  \\
    Others  &  0.002  &  0.007  &  0.077  &  0.059  &  0.005  &  0.065  \\
\hline\\[-0.3cm]
                Total  &  0.012  &  0.021  &  0.292  &  0.207  &  0.025  &  0.245  \\
\hline
\end{tabular*}
\vspace{1.5\baselineskip}

\vspace{-0.35cm}

\setlength{\tabcolsep}{0.0pc}
\begin{tabular*}{\textwidth}{@{\extracolsep{\fill}}lccccccc}
\hline
 \rule[-6pt]{0pt}{18pt}  &    $\IPzRe$  &    $\IPMIm$  &    $\IPMRe$  &      $\Uzm$  &      $\Uzp$  &   $\UMzmIm$  &   $\UMzmRe$  \\
\hline\\[-0.3cm]
 Dalitz plot model  &  0.017  &  0.007  &  0.127  &  0.082  &  0.041  &  0.144  &  0.209  \\
 $\rho$,$\rho'$,$\rho''$ lineshape  &  0.308  &  0.138  &  0.306  &  0.012  &  0.012  &  0.086  &  0.159  \\
   Fit bias  &  0.093  &  0.014  &  0.239  &  0.007  &  0.001  &  0.120  &  0.145  \\
 $N_{\rm Background}$  &  0.207  &  0.221  &  0.496  &  0.007  &  0.010  &  0.037  &  0.082  \\
    \B background \CP  &  0.121  &  0.168  &  0.088  &  0.015  &  0.005  &  0.052  &  0.044  \\
              Others  &  0.092  &  0.133  &  0.078  &  0.004  &  0.004  &  0.016  &  0.046  \\
\hline\\[-0.3cm]
                 Total  &  0.412  &  0.338  &  0.653  &  0.085  &  0.044  &  0.216  &  0.317  \\
\hline
\end{tabular*}
\vspace{1.5\baselineskip}

\vspace{-0.35cm}
\setlength{\tabcolsep}{0.0pc}
\begin{tabular*}{\textwidth}{@{\extracolsep{\fill}}lccccccc}
\hline
 \rule[-6pt]{0pt}{18pt}  &   $\UMzpIm$  &   $\UMzpRe$  &      $\UMm$  &      $\UMp$  &   $\UPzmIm$  &   $\UPzmRe$  &   $\UPzpIm$  \\
\hline\\[-0.3cm]
 Dalitz plot model  &  0.034  &  0.024  &  0.022  &  0.030  &  0.036  &  0.258  &  0.076  \\
 $\rho$,$\rho'$,$\rho''$ lineshape  &  0.222  &  0.045  &  0.010  &  0.030  &  0.050  &  0.216  &  0.089  \\
   Fit bias  &  0.010  &  0.023  &  0.001  &  0.007  &  0.214  &  0.038  &  0.001  \\
 $N_{\rm Background}$  &  0.038  &  0.051  &  0.020  &  0.009  &  0.080  &  0.073  &  0.051  \\
     \B background \CP  &  0.038  &  0.015  &  0.041  &  0.014  &  0.073  &  0.052  &  0.042  \\
               Others  &  0.011  &  0.007  &  0.015  &  0.010  &  0.038  &  0.037  &  0.032  \\
\hline\\[-0.3cm]
                  Total  &  0.232  &  0.078  &  0.054  &  0.048  &  0.250  &  0.353  &  0.138  \\
\hline
\end{tabular*}
\vspace{1.5\baselineskip}

\vspace{-0.35cm}
\setlength{\tabcolsep}{0.0pc}
\begin{tabular*}{\textwidth}{@{\extracolsep{\fill}}lcccccc}
\hline
 \rule[-6pt]{0pt}{18pt}  &   $\UPzpRe$  &   $\UPMmIm$  &   $\UPMmRe$  &   $\UPMpIm$  &   $\UPMpRe$  &      $\UPm$ \\
\hline\\[-0.3cm]
 Dalitz plot model  &  0.045  &  0.014  &  0.250  &  0.703  &  0.227  &  0.010 \\
 $\rho$,$\rho'$,$\rho''$ lineshape  &  0.140  &  0.169  &  0.200  &  0.169  &  0.159  &  0.031 \\
   Fit bias  &  0.003  &  0.130  &  0.001  &  0.010  &  0.035  &  0.007 \\
 $N_{\rm Background}$  &  0.106  &  0.288  &  0.114  &  0.099  &  0.083  &  0.016 \\
    \B background \CP  &  0.059  &  0.055  &  0.238  &  0.028  &  0.038  &  0.036 \\
               Others  &  0.060  &  0.045  &  0.112  &  0.012  &  0.079  &  0.009 \\
\hline\\[-0.3cm]
                  Total  &  0.199  &  0.366  &  0.430  &  0.730  &  0.305  &  0.053 \\
\hline
\end{tabular*}
\vspace{1.5\baselineskip}

\vspace{-0.35cm}
\setlength{\tabcolsep}{0.0pc}
\begin{tabular*}{\textwidth}{@{\extracolsep{\fill}}lcccccccc}
\hline
\rule[-6pt]{0pt}{18pt}  &   $\Acp$  &    $C$  &    $\dC$  &     $S$  &    $\dS$   & $\Czz$  & $\Szz$  & $\fzz$ \\
\hline\\[-0.3cm]
   Dalitz plot model  &  0.008  &  0.002  &  0.013  &  0.015  &  0.024  &  0.464  &  0.065  &  0.037 \\
 $\rho$,$\rho'$,$\rho''$ lineshape  &  0.011  &  0.021  &  0.011  &  0.016  &  0.008  &  0.048  &  0.010  &  0.007 \\
            Fit bias  &  0.015  &  0.026  &  0.081  &  0.024  &  0.055  &  0.236  &  0.062  &  0.002 \\
$N_{\rm Background}$  &   0.003  &   0.004  &   0.014  &   0.005  &   0.008    &   0.030  &   0.019  &   0.005 \\
    \B background \CP  &   0.005  &   0.032  &   0.006  &   0.018  &   0.007   &   0.059  &   0.014  &   0.002 \\
              Others  &   0.003  &   0.007  &   0.006  &   0.006  &   0.008  &   0.017  &   0.007  &   0.002 \\
\hline\\[-0.3cm]
                 Total  &   0.021  &   0.047  &   0.085  &   0.038  &   0.062  &   0.527  &   0.094  &   0.039 \\
\hline
\end{tabular*}

%% file: corrTableSyst1Vivace.tex
\setlength{\tabcolsep}{0.0pc}
\begin{tabular*}{\textwidth}{@{\extracolsep{\fill}}lccccccccccccc}
\hline
 && \\[-0.3cm]
\rule[-6pt]{0pt}{18pt}  & $\sigPiNb$  &      $\Iz$  &      $\IM$  &   $\IMzIm$  &   $\IMzRe$  &      $\IP$  &   $\IPzIm$  &   $\IPzRe$  &   $\IPMIm$  &   $\IPMRe$  &     $\Uzm$  &     $\Uzp$  &  $\UMzmIm$ \\[0.15cm]
\hline
 &&\\[-0.3cm] $\sigPiNb$  &  $\phantom{-}1.00$  &  &  &  &  &  &  &  &  &  &  &  & \\
      $\Iz$  &  $\phantom{-}0.10$  &  $\phantom{-}1.00$  &  &  &  &  &  &  &  &  &  &  & \\
      $\IM$  &  $\phantom{-}0.07$  &  $-0.20$  &  $\phantom{-}1.00$  &  &  &  &  &  &  &  &  &  & \\
   $\IMzIm$  &  $\phantom{-}0.00$  &  $-0.33$  &  $\phantom{-}0.78$  &  $\phantom{-}1.00$  &  &  &  &  &  &  &  &  & \\
   $\IMzRe$  &  $\phantom{-}0.29$  &  $\phantom{-}0.17$  &  $-0.52$  &  $-0.56$  &  $\phantom{-}1.00$  &  &  &  &  &  &  &  & \\
      $\IP$  &  $-0.13$  &  $\phantom{-}0.53$  &  $\phantom{-}0.04$  &  $-0.18$  &  $\phantom{-}0.31$  &  $\phantom{-}1.00$  &  &  &  &  &  &  & \\
   $\IPzIm$  &  $-0.07$  &  $-0.25$  &  $\phantom{-}0.58$  &  $\phantom{-}0.70$  &  $-0.25$  &  $\phantom{-}0.30$  &  $\phantom{-}1.00$  &  &  &  &  &  & \\
   $\IPzRe$  &  $\phantom{-}0.20$  &  $\phantom{-}0.36$  &  $-0.21$  &  $-0.10$  &  $-0.10$  &  $-0.51$  &  $-0.58$  &  $\phantom{-}1.00$  &  &  &  &  & \\
   $\IPMIm$  &  $-0.03$  &  $-0.35$  &  $-0.33$  &  $-0.14$  &  $\phantom{-}0.37$  &  $\phantom{-}0.23$  &  $\phantom{-}0.31$  &  $-0.55$  &  $\phantom{-}1.00$  &  &  &  & \\
   $\IPMRe$  &  $\phantom{-}0.36$  &  $\phantom{-}0.55$  &  $\phantom{-}0.01$  &  $\phantom{-}0.13$  &  $-0.20$  &  $-0.18$  &  $-0.18$  &  $\phantom{-}0.75$  &  $-0.54$  &  $\phantom{-}1.00$  &  &  & \\
     $\Uzm$  &  $\phantom{-}0.01$  &  $\phantom{-}0.90$  &  $-0.30$  &  $-0.39$  &  $\phantom{-}0.39$  &  $\phantom{-}0.71$  &  $-0.14$  &  $\phantom{-}0.09$  &  $-0.07$  &  $\phantom{-}0.28$  &  $\phantom{-}1.00$  &  & \\
     $\Uzp$  &  $-0.11$  &  $-0.75$  &  $\phantom{-}0.46$  &  $\phantom{-}0.58$  &  $-0.55$  &  $-0.71$  &  $\phantom{-}0.22$  &  $\phantom{-}0.04$  &  $-0.14$  &  $-0.12$  &  $-0.89$  &  $\phantom{-}1.00$  & \\
  $\UMzmIm$  &  $\phantom{-}0.49$  &  $-0.66$  &  $\phantom{-}0.29$  &  $\phantom{-}0.34$  &  $-0.11$  &  $-0.49$  &  $\phantom{-}0.29$  &  $-0.17$  &  $\phantom{-}0.25$  &  $-0.20$  &  $-0.73$  &  $\phantom{-}0.62$  &  $\phantom{-}1.00$ \\
  $\UMzmRe$  &  $-0.39$  &  $\phantom{-}0.53$  &  $-0.50$  &  $-0.55$  &  $\phantom{-}0.35$  &  $\phantom{-}0.55$  &  $-0.24$  &  $\phantom{-}0.03$  &  $\phantom{-}0.25$  &  $-0.08$  &  $\phantom{-}0.71$  &  $-0.71$  &  $-0.76$ \\
  $\UMzpIm$  &  $-0.07$  &  $\phantom{-}0.15$  &  $-0.40$  &  $-0.49$  &  $-0.22$  &  $-0.31$  &  $-0.54$  &  $\phantom{-}0.50$  &  $-0.19$  &  $\phantom{-}0.17$  &  $-0.00$  &  $-0.05$  &  $-0.11$ \\
  $\UMzpRe$  &  $-0.31$  &  $-0.58$  &  $\phantom{-}0.26$  &  $\phantom{-}0.12$  &  $-0.04$  &  $\phantom{-}0.08$  &  $\phantom{-}0.36$  &  $-0.75$  &  $\phantom{-}0.33$  &  $-0.84$  &  $-0.39$  &  $\phantom{-}0.26$  &  $\phantom{-}0.28$ \\
     $\UMm$  &  $-0.05$  &  $-0.44$  &  $-0.25$  &  $-0.22$  &  $\phantom{-}0.15$  &  $-0.26$  &  $-0.15$  &  $-0.10$  &  $\phantom{-}0.46$  &  $-0.34$  &  $-0.38$  &  $\phantom{-}0.24$  &  $\phantom{-}0.36$ \\
     $\UMp$  &  $-0.10$  &  $\phantom{-}0.76$  &  $-0.24$  &  $-0.35$  &  $\phantom{-}0.04$  &  $\phantom{-}0.36$  &  $-0.26$  &  $\phantom{-}0.39$  &  $-0.24$  &  $\phantom{-}0.33$  &  $\phantom{-}0.74$  &  $-0.61$  &  $-0.67$ \\
  $\UPzmIm$  &  $\phantom{-}0.87$  &  $\phantom{-}0.07$  &  $-0.10$  &  $-0.17$  &  $\phantom{-}0.51$  &  $-0.07$  &  $-0.16$  &  $\phantom{-}0.14$  &  $\phantom{-}0.17$  &  $\phantom{-}0.10$  &  $\phantom{-}0.06$  &  $-0.20$  &  $\phantom{-}0.48$ \\
  $\UPzmRe$  &  $-0.18$  &  $-0.50$  &  $-0.00$  &  $\phantom{-}0.00$  &  $-0.41$  &  $-0.74$  &  $-0.38$  &  $\phantom{-}0.35$  &  $-0.14$  &  $-0.12$  &  $-0.62$  &  $\phantom{-}0.59$  &  $\phantom{-}0.24$ \\
  $\UPzpIm$  &  $-0.05$  &  $-0.50$  &  $-0.09$  &  $-0.01$  &  $-0.19$  &  $-0.27$  &  $\phantom{-}0.16$  &  $-0.27$  &  $\phantom{-}0.18$  &  $-0.18$  &  $-0.59$  &  $\phantom{-}0.40$  &  $\phantom{-}0.52$ \\
  $\UPzpRe$  &  $\phantom{-}0.03$  &  $-0.23$  &  $-0.06$  &  $-0.08$  &  $\phantom{-}0.52$  &  $\phantom{-}0.49$  &  $\phantom{-}0.35$  &  $-0.79$  &  $\phantom{-}0.65$  &  $-0.58$  &  $\phantom{-}0.05$  &  $-0.25$  &  $\phantom{-}0.14$ \\
  $\UPMmIm$  &  $\phantom{-}0.33$  &  $\phantom{-}0.20$  &  $\phantom{-}0.27$  &  $\phantom{-}0.47$  &  $-0.27$  &  $-0.12$  &  $\phantom{-}0.28$  &  $\phantom{-}0.25$  &  $-0.34$  &  $\phantom{-}0.65$  &  $-0.05$  &  $\phantom{-}0.17$  &  $\phantom{-}0.24$ \\
  $\UPMmRe$  &  $\phantom{-}0.05$  &  $\phantom{-}0.62$  &  $\phantom{-}0.25$  &  $\phantom{-}0.22$  &  $-0.05$  &  $\phantom{-}0.35$  &  $\phantom{-}0.14$  &  $\phantom{-}0.26$  &  $-0.43$  &  $\phantom{-}0.43$  &  $\phantom{-}0.61$  &  $-0.38$  &  $-0.55$ \\
  $\UPMpIm$  &  $-0.08$  &  $-0.84$  &  $\phantom{-}0.27$  &  $\phantom{-}0.34$  &  $-0.50$  &  $-0.69$  &  $\phantom{-}0.14$  &  $-0.09$  &  $\phantom{-}0.12$  &  $-0.33$  &  $-0.94$  &  $\phantom{-}0.86$  &  $\phantom{-}0.72$ \\
  $\UPMpRe$  &  $-0.04$  &  $\phantom{-}0.62$  &  $\phantom{-}0.13$  &  $\phantom{-}0.09$  &  $\phantom{-}0.21$  &  $\phantom{-}0.68$  &  $\phantom{-}0.27$  &  $-0.20$  &  $-0.22$  &  $\phantom{-}0.17$  &  $\phantom{-}0.69$  &  $-0.54$  &  $-0.49$ \\
     $\UPm$  &  $\phantom{-}0.14$  &  $\phantom{-}0.39$  &  $-0.34$  &  $-0.34$  &  $\phantom{-}0.10$  &  $-0.06$  &  $-0.40$  &  $\phantom{-}0.53$  &  $-0.05$  &  $\phantom{-}0.36$  &  $\phantom{-}0.32$  &  $-0.25$  &  $-0.15$ \\
[0.15cm]\hline
\end{tabular*}

%% file: corrTableSyst2Vivace.tex
\setlength{\tabcolsep}{0.0pc}
\begin{tabular*}{\textwidth}{@{\extracolsep{\fill}}lcccccccccccccc}
\hline
 && \\[-0.3cm]
\rule[-6pt]{0pt}{18pt}  &  $\UMzmRe$  &  $\UMzpIm$  &  $\UMzpRe$  &     $\UMm$  &     $\UMp$  &  $\UPzmIm$  &  $\UPzmRe$  &  $\UPzpIm$  &  $\UPzpRe$  &  $\UPMmIm$  &  $\UPMmRe$  &  $\UPMpIm$  &  $\UPMpRe$  &     $\UPm$ \\[0.15cm]
\hline
 &&\\[-0.3cm]  $\UMzmRe$  &  $\phantom{-}1.00$  &  &  &  &  &  &  &  &  &  &  &  &  & \\
  $\UMzpIm$  &  $\phantom{-}0.29$  &  $\phantom{-}1.00$  &  &  &  &  &  &  &  &  &  &  &  & \\
  $\UMzpRe$  &  $-0.14$  &  $-0.29$  &  $\phantom{-}1.00$  &  &  &  &  &  &  &  &  &  &  & \\
     $\UMm$  &  $-0.04$  &  $\phantom{-}0.11$  &  $\phantom{-}0.16$  &  $\phantom{-}1.00$  &  &  &  &  &  &  &  &  &  & \\
     $\UMp$  &  $\phantom{-}0.73$  &  $\phantom{-}0.51$  &  $-0.40$  &  $-0.42$  &  $\phantom{-}1.00$  &  &  &  &  &  &  &  &  & \\
  $\UPzmIm$  &  $-0.17$  &  $-0.05$  &  $-0.14$  &  $\phantom{-}0.14$  &  $-0.07$  &  $\phantom{-}1.00$  &  &  &  &  &  &  &  & \\
  $\UPzmRe$  &  $-0.14$  &  $\phantom{-}0.58$  &  $\phantom{-}0.15$  &  $\phantom{-}0.37$  &  $-0.09$  &  $-0.12$  &  $\phantom{-}1.00$  &  &  &  &  &  &  & \\
  $\UPzpIm$  &  $-0.47$  &  $\phantom{-}0.14$  &  $\phantom{-}0.19$  &  $\phantom{-}0.28$  &  $-0.57$  &  $-0.13$  &  $\phantom{-}0.11$  &  $\phantom{-}1.00$  &  &  &  &  &  & \\
  $\UPzpRe$  &  $\phantom{-}0.04$  &  $-0.67$  &  $\phantom{-}0.53$  &  $\phantom{-}0.22$  &  $-0.46$  &  $\phantom{-}0.20$  &  $-0.53$  &  $\phantom{-}0.14$  &  $\phantom{-}1.00$  &  &  &  &  & \\
  $\UPMmIm$  &  $-0.55$  &  $-0.29$  &  $-0.46$  &  $-0.28$  &  $-0.23$  &  $\phantom{-}0.07$  &  $-0.39$  &  $\phantom{-}0.28$  &  $-0.16$  &  $\phantom{-}1.00$  &  &  &  & \\
  $\UPMmRe$  &  $\phantom{-}0.30$  &  $-0.10$  &  $-0.34$  &  $-0.73$  &  $\phantom{-}0.64$  &  $-0.09$  &  $-0.36$  &  $-0.69$  &  $-0.32$  &  $\phantom{-}0.12$  &  $\phantom{-}1.00$  &  &  & \\
  $\UPMpIm$  &  $-0.60$  &  $\phantom{-}0.20$  &  $\phantom{-}0.42$  &  $\phantom{-}0.42$  &  $-0.59$  &  $-0.09$  &  $\phantom{-}0.73$  &  $\phantom{-}0.58$  &  $-0.16$  &  $-0.05$  &  $-0.62$  &  $\phantom{-}1.00$  &  & \\
  $\UPMpRe$  &  $\phantom{-}0.21$  &  $-0.53$  &  $-0.07$  &  $-0.55$  &  $\phantom{-}0.29$  &  $-0.08$  &  $-0.81$  &  $-0.39$  &  $\phantom{-}0.28$  &  $\phantom{-}0.31$  &  $\phantom{-}0.59$  &  $-0.77$  &  $\phantom{-}1.00$  & \\
     $\UPm$  &  $\phantom{-}0.35$  &  $\phantom{-}0.46$  &  $-0.48$  &  $\phantom{-}0.38$  &  $\phantom{-}0.45$  &  $\phantom{-}0.24$  &  $\phantom{-}0.17$  &  $-0.23$  &  $-0.36$  &  $-0.03$  &  $-0.02$  &  $-0.21$  &  $-0.14$  &  $\phantom{-}1.00$ \\
[0.15cm]\hline
\end{tabular*}

%% file: PhysicsResults.tex
\section{INTERPRETATION OF THE RESULTS}
\label{sec:Physics}

We can use the results of this time-dependent Dalitz analysis to 
extract the $\Bz(\Bzb) \to \rho^\pm \pi^\mp$ parameters defined in  Ref.~\cite{rhopipaper}:
\beqn
\label{eq:thTime}
  \lefteqn{f^{\rho^\pm \pi^\mp}_{Q_{\rm tag}}(\deltat) = (1\pm \Acp)
           \frac{e^{-\left|\deltat\right|/\tau}}{4\tau}} \\
        &&\hspace{1.3cm}\times\,\bigg[1+Q_{\rm tag} 
             (S \pm \dS )\sin(\deltamd\deltat)\nonumber\\[-0.1cm]
        &&\hspace{1.3cm}\phantom{\times\,\bigg[1}
            -Q_{\rm tag} 
                (C \pm \dC)\cos(\deltamd\deltat)\bigg]\;,\nonumber
\eeqn
where $Q_{\rm tag}= 1(-1)$ when the tagging meson $\Bz_{\rm tag}$
is a $\Bz(\Bzb)$.
The time- and flavor-integrated charge asymmetry $\Acp$
measures direct \CP violation and the quantities $S$ and $C$ 
parameterize mixing-induced \CP violation related to the angle $\alpha$,
and flavor-dependent direct \CP violation, respectively.
The parameters $\dC_{\rho \pi}$ and $\dS_{\rho\pi}$ are insensitive to
\CP violation.

The $U$ and $I$ coefficients are related to the 
 parameters as follows:
\beqn
\label{eq:q2bparams}
	C^+ = \frac{ U^-_+ }{ U^+_+ }~, &	& C^- = \frac{ U^-_- }{ U^+_- }~, \nonumber\\
	S^+ = \frac{ 2 \, I_+ }{ U^+_+ }~,&&S^- = \frac{ 2 \, I_- }{ U^+_- }~,\nonumber\\
	\Acp = \frac{ U^+_+ \, - U^+_- }{ U^+_+ \, + U^+_- }~,&&
\eeqn
%\end{widetext}
 where $C=(C^++C^-)/2$, $\dC=(C^+-C^-)/2$, $S=(S^++S^-)/2$, 
and $\dS=(S^+-S^-)/2$ . The definitions of Eq.~(\ref{eq:q2bparams})
explicitly account for the presence of interference effects, and are 
thus exact even for a $\rho$ with finite width, as long as the $U$ and 
$I$ coefficients are obtained with a Dalitz plot analysis. This treatment
leads to  slightly increased statistical uncertainties compared to 
the results obtained neglecting the interference effects.

Using a least-squares method including statistical and systematic correlations for the $U$ and $I$ coefficients, we obtain: 
\beqn
	\Acp     &=&     -0.14\pm 0.05 \pm{0.02}~, \\
	C   	&=& \ph{-}0.15 \pm 0.09 \pm 0.05~, \\
	S 	&=&  -0.03\pm 0.11\pm 0.04~,
\eeqn
where the first errors  are statistical and the second 
are the systematic uncertainties.
For the other parameters in the description of the 
$\Bz(\Bzb) \to \rhopi$ decay-time dependence, we measure
\beqn
	\dC 	&=& \phantom{-}0.39\pm 0.09\pm 0.09~, \\
	\dS 	&=& -0.01\pm 0.14\pm 0.06~.
\eeqn
In addition, we measure the $B^0\to\rho^0\pi^0$ \CP-violation parameters and decay fraction to be
\beqn
	C_{00}  =\frac{U_0^-}{U_0^+}=&-0.10 \pm 0.40 \pm 0.53~, \\
	S_{00}	=\frac{2 \, I_0}{U_0^+}=& \ph{-}0.04 \pm 0.44\pm 0.18~,\\
	f_{00}	=\frac{U_0^+}{U_+^+ + U_-^+ + U_0^+}=& \ph{-}0.136 \pm 0.036\pm 0.039~.
\eeqn
The systematic errors are dominated by the 
uncertainty on the \CP content of the \B-related backgrounds.
Other contributions are the signal description in the likelihood 
model (including the limit on non-resonant $\Btopipipi$ events), and
the fit bias uncertainty. The large systematic error on $C_{00}$ is 
due to the possible $\pi^+\pi^-$ S-wave contribution. 
The correlation matrix, including statistical and 
systematic uncertainties, of the eight quasi-two-body parameters is given
in Table~\ref{tab:q2bCov}.

\begin{table*}[t]
\begin{center}
\caption{ \label{tab:q2bCov}
        Correlation matrix  of the quasi-two-body parameters.}
\input{corrTableQ2B.tex}
\end{center}
\end{table*}

One can transform the experimentally convenient (uncorrelated)
 direct-\CP violation parameters $\Crhopi$ and $\Acp$
into  $\Acppm$, $\Acpmp$,
defined by
\beqn
\label{eq:Adirpm}
    \Acppm &=& \frac{|\kappm|^2-1}{|\kappm|^2+1}
    \;=\; -\frac{\Acp+\Crhopi+\Acp\dCrhopi}{1+\dCrhopi+\Acp\Crhopi}
    ~,\\[0.2cm]\nonumber
\label{eq:Adirmp}
    \Acpmp &=& \frac{|\kapmp|^2-1}{|\kapmp|^2+1}
    \;=\; \frac{\Acp-\Crhopi-\Acp\dCrhopi}{1-\dCrhopi-\Acp\Crhopi}
    ~,
\eeqn
where
$\kappm = (q/p)\Ampb/\Apm$ and $\kapmp =(q/p)\Apmb/\Amp$,
so that $\Acppm$ ($\Acpmp$) involves only diagrams where the $\rho$
($\pi$) meson is formed from the $W$ boson. We find
\begin{figure}[t]
  \centerline{  \epsfxsize7.7cm\epsffile{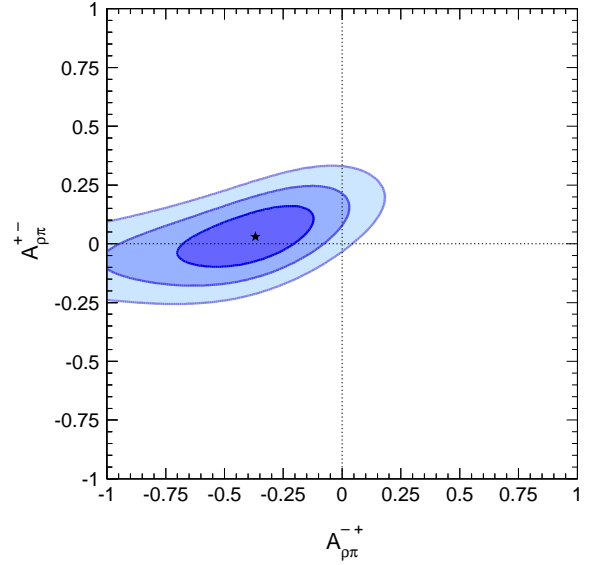}}
  \caption{\label{fig:apmamp} 
	Confidence level contours for the direct \CP asymmetries
	$\Acppm$ versus $\Acpmp$. The shaded areas represent 
	$1\sigma$, $2\sigma$ and $3\sigma$ contours, respectively. }
\end{figure}
\beqn
    \Acppm &=& \phantom{-}0.03\pm0.07\pm0.04~, \\
    \Acpmp &=& -0.37\pm0.16^{\,+0.09}_{\,-0.10}~,
\eeqn
with a correlation coefficient of $0.62$ between $\Acppm$
and $\Acpmp$. The confidence level contours including systematic
uncertainties are shown in Fig.~\ref{fig:apmamp}. The significance, including 
systematic uncertainties and calculated by using a minimum $\chi^2$
method, of  direct \CP violation is 
less than $3.0\sigma$. 
\begin{figure*}[t]
  \centerline{  \epsfxsize8.2cm\epsffile{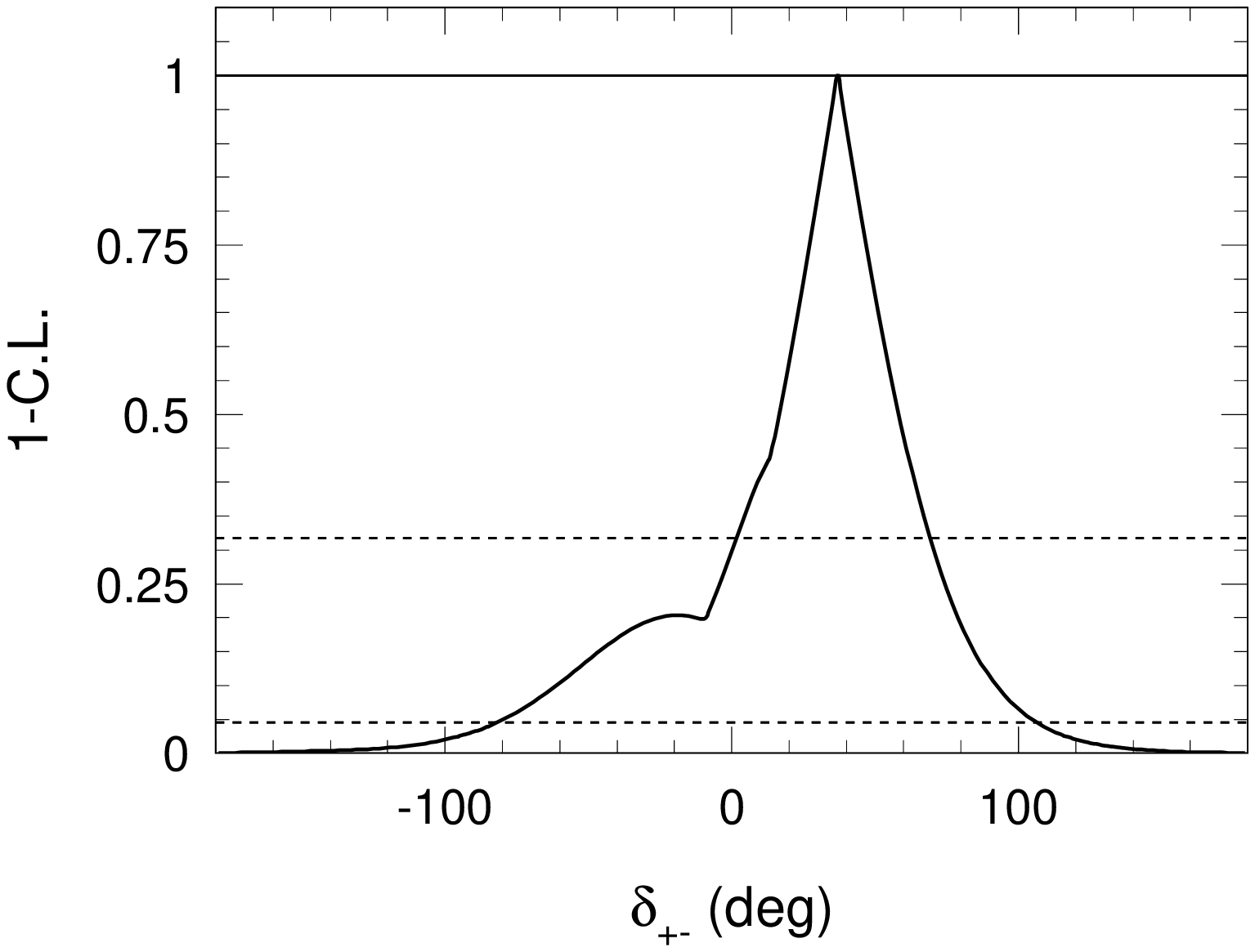}
   	        \epsfxsize8.2cm\epsffile{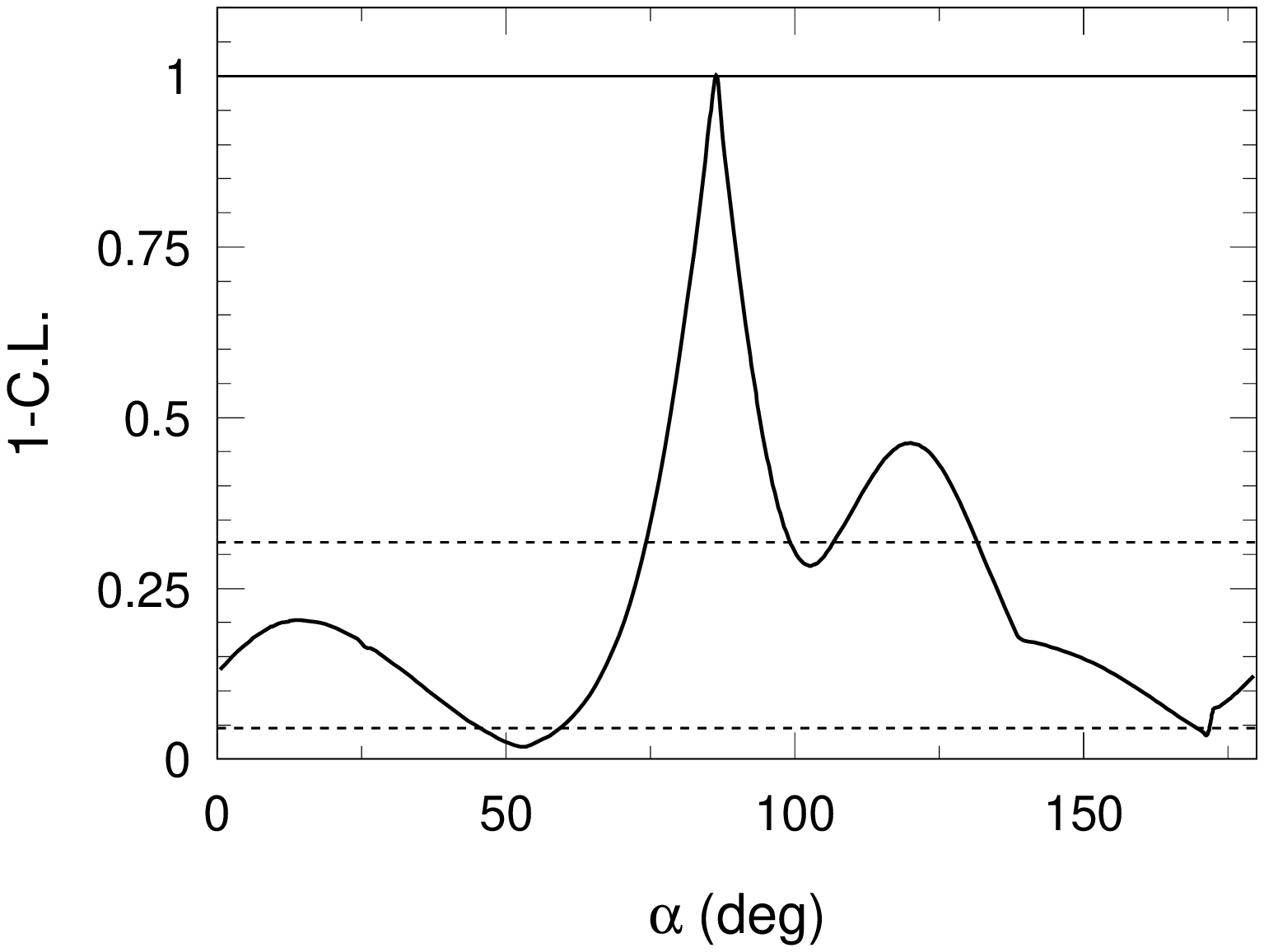}}
  \caption{\label{fig:deltaalpha} 
	Confidence level functions for $\delta_{+-}$ (left) and 
	$\alpha$ (right). Indicated 
	by the dashed horizontal lines are the confidence level (C.L.) values
        corresponding to $1\sigma$ and $2\sigma$, respectively.}
\end{figure*}

The measurement of the resonance interference terms allows us to 
determine the relative phase
\beq
\label{eq:deltapm}
\delta_{+-} = \arg\left( A^{+*}A^{-} \right)~ 
\eeq
between the amplitudes of the 
decays $B^0\to\rho^-\pi^+$ and $B^0\to\rho^+\pi^-$. Through 
the definitions in Eqs. (\ref{eq:firstObs})--(\ref{eq:lastObs}), we can derive 
a constraint on $\delta_{+-}$ from the measured $U$ and $I$ 
coefficients
by performing a least-squares minimization with the six complex amplitudes
as free parameters. 
The constraint can be improved with the use of strong 
isospin symmetry. The amplitudes 
$\Aij$ represent the sum of tree-level and penguin-type
amplitudes, which have different CKM factors: the tree-level $(\Tij)$
$\Bz\to\rho^\kappa\pi^{\kappab}$ transition amplitude
is proportional to 
$V_{ud}V_{ub}^*$, while the corresponding penguin-type amplitude $(\Pij)$
involves $V_{qd}V_{qb}^*$, where $q=u,c,t$. 
Here we denote by $\kappab$ the charge conjugate of $\kappa$,
where $\overline \kappa=0$ when $\kappa=0$. 
Using the unitarity of the 
CKM matrix one can reorganize the amplitudes and obtain~\cite{BaBarPhysBook}
\beqn
\label{eq:aijamps}
	\Aij 		&=& \Tij e^{-i\alpha} + \Pij ~, \nonumber\\
	(q/p)\Abij 		&=& \Tji e^{+i\alpha} + \Pji ~,
\eeqn
where the magnitudes of the CKM factors have been absorbed in 
$\Tij$, $\Pij$, $\Tji$ and $\Pji$.  
The Eqs.~(\ref{eq:aijamps}) represent 13 unknowns
of which two can be fixed due to an arbitrary
global phase and the normalization condition $U_+^+=1$. Using
strong isospin symmetry 
one can identify $P^{0}=-(P^{+}+P^{-})/2$, which reduces the 
number of unknowns to be determined by the fit to nine. This set of parameters 
provides the constraint on $\delta_{+-}$, shown in the 
left plot of Fig.~\ref{fig:deltaalpha}. 
We find for the solution that is favored by the fit
\beq
	\delta_{+-} \; = \; \left(37\,\pm37\right)^\circ~,
\eeq
where the errors include both statistical and systematic effects.
There is  only a marginal constraint on $\delta_{+-}$  obtained at 95\% confidence level (C.L.).

Finally, following the same procedure, we can also
derive a constraint on $\alpha$ from the measured $U$ and $I$ 
coefficients. The resulting
C.L. function versus $\alpha$ is given in the right-hand plot
of Fig.~\ref{fig:deltaalpha},  including systematic uncertainties.
Ignoring the mirror solution at $\alpha + 180^\circ$, we find at $68\%$ C.L. 
\beq
	\alpha \; = \; \left(87\,^{+45}_{-13}\right)^\circ~.
\eeq
Almost no constraint on $\alpha$ is achieved at two sigma and beyond.

%% file: corrTableQ2B.tex
\setlength{\tabcolsep}{0.0pc}
\begin{tabular*}{\textwidth}{@{\extracolsep{\fill}}lcccccccc}
\hline
 && \\[-0.3cm]
\rule[-6pt]{0pt}{18pt}  &     $\Acp$  &       $\C$  &      $\dC$  &       $\S$  &      $\dS$  &     $\Czz$  &     $\Szz$  &     $\fzz$ \\[0.15cm]
\hline
 &&\\[-0.3cm]     $\Acp$  &  $\phantom{-}1.00$  &  &  &  &  &  &  & \\
       $\C$  &  $-0.06$  &  $\phantom{-}1.00$  &  &  &  &  &  & \\
      $\dC$  &  $\phantom{-}0.12$  &  $\phantom{-}0.38$  &  $\phantom{-}1.00$  &  &  &  &  & \\
       $\S$  &  $-0.07$  &  $-0.13$  &  $-0.15$  &  $\phantom{-}1.00$  &  &  &  & \\
      $\dS$  &  $-0.04$  &  $-0.12$  &  $-0.26$  &  $\phantom{-}0.33$  &  $\phantom{-}1.00$  &  &  & \\
     $\C_{00}$  &  $-0.38$  &  $-0.15$  &  $-0.18$  &  $\phantom{-}0.19$  &  $\phantom{-}0.22$  &  $\phantom{-}1.00$  &  & \\
     $\S_{00}$  &  $-0.24$  &  $\phantom{-}0.18$  &  $\phantom{-}0.50$  &  $-0.18$  &  $-0.14$  &  $\phantom{-}0.35$  &  $\phantom{-}1.00$  & \\
     $f_{00}$  &  $\phantom{-}0.19$  &  $-0.04$  &  $-0.09$  &  $-0.17$  &  $-0.38$  &  $-0.84$  &  $-0.70$  &  $\phantom{-}1.00$ \\
[0.15cm]\hline
\end{tabular*}

%% file: Summary.tex
\section{SUMMARY}
\label{sec:Summary}

We have presented a measurement of 
\CP-violating asymmetries in $\Btopipipi$ decays dominated by 
the $\rho$ resonance. The results are obtained from a data sample 
of 375 million $\FourS \to B\Bbar$ decays. We perform a time-dependent 
Dalitz plot analysis. From the measurement of the coefficients of 26 form-factor 
bilinears we determine the three \CP-violating 
and two \CP-conserving quasi-two-body parameters, and  find no evidence of direct \CP violation.
 Taking advantage of 
the interference between the $\rho$ resonances in the Dalitz plot,
we derive constraints on the relative strong phase between 
$\Bz$ decays to $\rho^+\pim$ and $\rho^-\pip$, and on the angle
$\alpha$ of the Unitarity Triangle. These measurements are 
consistent with the results obtained by Belle \cite{BELLErhopiDalitz} 
as well as with the expectation of a SM fit to all constraints on the CKM matrix~\cite{alphaSM,UTFit}.

%% file: pubboard/acknowledgements.tex
We are grateful for the 
extraordinary contributions of our \pep2\ colleagues in
achieving the excellent luminosity and machine conditions
that have made this work possible.
The success of this project also relies critically on the 
expertise and dedication of the computing organizations that 
support \babar.
The collaborating institutions wish to thank 
SLAC for its support and the kind hospitality extended to them. 
This work is supported by the
US Department of Energy
and National Science Foundation, the
Natural Sciences and Engineering Research Council (Canada),
Institute of High Energy Physics (China), the
Commissariat \`a l'Energie Atomique and
Institut National de Physique Nucl\'eaire et de Physique des Particules
(France), the
Bundesministerium f\"ur Bildung und Forschung and
Deutsche Forschungsgemeinschaft
(Germany), the
Istituto Nazionale di Fisica Nucleare (Italy),
the Foundation for Fundamental Research on Matter (The Netherlands),
the Research Council of Norway, the
Ministry of Science and Technology of the Russian Federation, 
Ministerio de Educaci\'on y Ciencia (Spain), and the
Particle Physics and Astronomy Research Council (United Kingdom). 
Individuals have received support from 
the Marie-Curie IEF program (European Union) and
the A. P. Sloan Foundation.

%% file: prd.bbl
\begin{thebibliography}{99}

\bibitem{BabarS2b}	\babar\ Collaboration, B.~Aubert {\em et al.}, 
			\jprl{94}, 161803 (2005).	

\bibitem{BelleSin2beta}	Belle Collaboration (K. F. Chen\ea), 
			\jprl{98}, 031802 (2007). 

\bibitem{CKM}      	N.~Cabibbo, 
                        Phys. Rev. Lett. {\bf 10}, 531 (1963);
			M.~Kobayashi and T.~Maskawa, 
                        Prog. Theor. Phys. {\bf 49}, 652 (1973).

\bibitem{ccref}             Charge conjugate decay modes are 
                                assumed unless explicity stated.

\bibitem{babarpipi}	\babar\ Collaboration, (B.~Aubert {\em et al.}),
                      	\jprl{95}, 151803 (2005).
			
\bibitem{bellepipi}	Belle Collaboration (K.~Abe \ea),
			\jprl{95}, 10181 (2005).

\bibitem{babarrhorho}	\babar\ Collaboration (B.~Aubert \ea), 
			\jprl{95}, 041805 (2005).

\bibitem{bellerhorho}	Belle Collaboration (K.~Abe \ea), 
			\jprl{96}, 171801 (2006).

\bibitem{GLisospin}	M.~Gronau and D.~London, 
                        \jprl{65}, 3381 (1990).

\bibitem{Lipkinetal} 	H.J.~Lipkin, Y.~Nir, H.R.~Quinn and A.~Snyder,
                        Phys. Rev. {\bf D 44}, 1454 (1991).

\bibitem{SnyderQuinn}   H.R.~Quinn and A.E.~Snyder, 
                        Phys. Rev. {\bf D 48}, 2139 (1993).

\bibitem{PDG}		Particle Data Group (S.~Eidelman {\em et al.}), 
			Phys. Lett. {\bf B 592}, 1 (2004).

\bibitem{taueeref}	ALEPH Collaboration, (R.~Barate \ea),
			Z. Phys. {\bf C 76}, 15 (1997); we use updated 
			lineshape fits including new data from $\epem$
			annihilation~\cite{cmd2} and $\tau$ spectral
			functions~\cite{aleph_new} (masses and widths
			in \mevcc):
			$m_{\rho^\pm(770)} = 775.5 \pm 0.6$, 
			$m_{\rho^0(770)}=773.1 \pm 0.5 $,
			$\Gamma_{\rho^\pm(770)}=148.2 \pm 0.8$,
			$\Gamma_{\rho^0(770)}=148.0 \pm 0.9 $,
			$m_{\rho(1450)}=1409  \pm 12$,
			$\Gamma_{\rho(1450)}= 500  \pm 37$,
			$m_{\rho(1700)}=1749  \pm 20$,
			and $\Gamma_{\rho(1700)}\equiv235$.

\bibitem{BaBarPhysBook}	The \babar\ Physics Book, 
			Editors P.F.~Harrison and H.R.~Quinn, 
			SLAC-R-504 (1998).

\bibitem{rhoGS}		G.J.~Gounaris and J.J.~Sakurai,
                        Phys. Rev. Lett. {\bf 21}, 244 (1968).

\bibitem{quinnsilva}    H.R.~Quinn and J.~Silva,
                      	Phys. Rev. \textbf{D 62}, 054002 (2000).

\bibitem{rhopipaper}    \babar\   Collaboration (B.~Aubert \ea),
                        Phys. Rev. Lett. {\bf 91}, 201802 (2003).
%                        updated preliminary results at \babar-PLOT-0055 (2003).

\bibitem{geant}	GEANT4 Collaboration (S.~Agostinelli \ea), Nucl. Instrum. Methods {\bf A 506}, 250 (2003).

\bibitem{babarNim}	\babar\ Collaboration, 
			B.~Aubert {\em et al.}, 
			Nucl. Instrum. Methods {\bf A 479}, 1 (2002).

\bibitem{NNo}           P.~Gay, B.~Michel, J.~Proriol, and O.~Deschamps,
                        ``{\em Tagging Higgs Bosons in Hadronic LEP-2
                        Events with Neural Networks.}'',
                        In Pisa 1995, New computing techniques in 
                        physics research, 725 (1995).


\bibitem{HFAG}		Heavy Flavor Averaging Group (K.~Anikeev {\em et al.}),
			May (2005), hep-ex/0505100. 

\bibitem{PDFsCB}	T. Skwarnicki, DESY F31-86-02, Ph.D. thesis (1986);
			see also Ref.~\cite{BabarS2b}.
			
\bibitem{keys}
			  K.~S.~Cranmer,
			  Comput.\ Phys.\ Commun.\  {\bf 136}, 198 (2001).

\bibitem{PDFsArgus}	ARGUS Collaboration (H. Albrecht \ea), 
			Z. Phys. C {\bf 48}, 543 (1990).

\bibitem{BABARrho0pi0}	\babar\   Collaboration (B.~Aubert \ea),
			Phys. Rev. Lett. {\bf 93}, 051802 (2004).

\bibitem{BELLErho0pi0}	Belle Collaboration (J.~Dragic \ea),
                        Phys. Rev. {\bf D 73}, 111105 (2006).

\bibitem{BELLErhopiDalitz} 	Belle Collaboration (A.~Kusaka \ea), submitted to PRL, hep-ex/0701015.

\bibitem{alphaSM}	J.~Charles \ea, 
			Eur. Phys. J. {\bf C 41}, 1 (2005).

\bibitem{UTFit} 	M.~Bona {\it et al.},
		  JHEP {\bf 0507}, 028 (2005).	



\bibitem{cmd2}        	CMD-2 Collaboration (R.R. Akhmetshin {\it et al.}),
                      	 Phys. Lett. {\bf B 527}, 161 (2002).


\bibitem{aleph_new}   	ALEPH Collaboration ( S.~Schael {\it et al.} ), 
			Phys.\ Rep.,  {\bf 421}, 191 (2005).
\end{thebibliography}
